\documentclass[aps,prd,onecolumn,nofootinbib,superscriptaddress,floatfix]{revtex4-2}
\usepackage{graphicx}
\usepackage[hidelinks]{hyperref}
\usepackage{amsmath}
\usepackage{amssymb}
\usepackage{booktabs}

\newcommand{\MPl}{\ensuremath{M_{\mathrm{P}}}}
\newcommand{\lPl}{\ensuremath{\ell_{\mathrm{P}}}}

\newcommand{\Mmin}{\ensuremath{M_{\min}}}

\newcommand{\Tmax}{\ensuremath{T_H^{\max}}}
\newcommand{\Srem}{\ensuremath{S_{\mathrm{rem}}}}

\begin{document}
\title{Thermodynamics and interior structure of density-responsive
regular black holes}

\author{Martin Drobczyk}
\email[]{martin.drobczyk@dlr.de}
\affiliation{Institute of Space Systems,
German Aerospace Center (DLR), Bremen, Germany}

\begin{abstract}
This work studies regular black holes motivated by the scalar sector of density-responsive gravity (DRG). An algebraic prescription with phenomenological local running maps a Gaussian matter profile to a diagnostic density profile. A pressure check shows that this profile does not solve the full static DRG equations. It serves only to select an effective metric family with two horizons. The extremal mass is calibrated to $\Mmin = \sqrt{1+A/2}\,\MPl \simeq 1.006\,\MPl$, where $A=0.024$ is the historical scalar response amplitude. A fit to $\rho(r) = \rho_0\exp[-(r/r_c)^n]$ selects $n=2$. The temperature is nonmonotonic, reaching $\Tmax=0.02547\,\MPl$ at $M=1.332\,\MPl$ before vanishing at $\Mmin$. The calculation verifies the generic scaling near extremality $\mathcal{Z}_{\rm tr}\propto(M-\Mmin)^{-1/2}$ of a static lapse diagnostic for $n=2,3,4,6$ and for the Hayward and Bardeen controls. A regulated integral of coordinate time between the horizons provides a static cavity diagnostic, not a prediction of echoes of gravitational waves. The remaining area entropy equals the area entropy already lost at $M_{\rm cross}/M_0=0.707107\simeq1/\sqrt2$. This geometric crossing does not establish a Page curve or unitarity. The endpoint area entropy is below $2.6\%$ of the initial Bekenstein-Hawking entropy for $M_0 = 5\,\MPl$ and below $0.03\%$ for $M_0 = 50\,\MPl$. An inverse source reconstruction confirms an isotropic core of de Sitter type but retains the pressure mismatch. An exploratory collapse calculation with an imposed bounce factor produces bounded configurations with $1-2m/r<0$ without establishing formation of the calibrated geometry. The robust results are the calibrated thermodynamics and generic scaling near extremality.
\end{abstract}


\maketitle

\section{Introduction}
\label{sec:intro}

Hawking evaporation connects the spacetime geometry of black holes to quantum radiation and thermodynamics in general relativity (GR). Semiclassical arguments predict that black holes radiate thermally through the Hawking process, leading to gradual mass loss and eventual evaporation \cite{Hawking1975}. If taken at face value, this process maps an initially pure quantum state to a mixed final state, giving rise to the black hole information paradox \cite{Hawking1976}. Despite decades of progress, a fully consistent description of black hole evaporation that respects both quantum unitarity and the equivalence principle remains an open problem.

A central obstacle is the presence of spacetime singularities in classical general relativity. Singularities signal a breakdown of the classical description and obscure the fate of infalling information. Any consistent resolution is widely expected to involve a modification of the deep interior, while leaving the near-horizon region smooth and semiclassical, avoiding departures from classical geometry such as firewalls \cite{Almheiri2013}.

Regular black holes, spacetimes with a core of de~Sitter type replacing the central singularity, have been studied in phenomenological settings and in settings inspired by quantum gravity. Bardeen \cite{Bardeen1968}, Hayward \cite{Hayward2005} and Dymnikova \cite{Dymnikova1992} introduced effective mass functions that regularize the central region. These constructions share universal properties near extremality (a minimum mass, a maximum temperature and a degenerate horizon in the extremal limit \cite{Frolov2016}) but typically postulate the interior geometry rather than deriving it from an underlying theory. Within quantum gravity, regular cores and remnants arise through improvement using the renormalization group in asymptotic safety \cite{Bonanno2000} or through bounce mechanisms in loop quantum gravity \cite{Rovelli2014}.  In asymptotic safety, Bonanno and Reuter~\cite{Bonanno2000} obtain remnant masses of order $\MPl$ with a core scale set by the gravitational antiscreening length. In loop quantum gravity, the Planck star of Rovelli and Vidotto~\cite{Rovelli2014} has a core determined by the minimum-area gap. In both cases the ultraviolet scale is an independent parameter of quantum gravity.

The density-responsive gravity (DRG) framework \cite{Drobczyk2025_1, Drobczyk2025_2} motivates the algebraic density response tested here. Its conformal prescription relates the physical metric to an auxiliary metric through $\tilde{g}_{\mu\nu} = \Omega^2(\rho_m)\, g_{\mu\nu}$, where the conformal factor $\Omega$ depends on the matter density $\rho_m$. This relation alone supplies neither a complete source action nor a derivation of a regular black hole. The amplitude $A = 0.024$ of the scalar response and the infrared mass scale $M_U \simeq 5.84\,\mathrm{meV}$ are retained as historical reference inputs, not as a cosmological determination of the core. The radial metric family and its extremal normalization are choices of the effective model calibrated below. The specific contribution is the comparison of the density prescription with that family, including a pressure consistency check that tests whether the prescribed source can support the metric.

The analysis tests how the DRG scalar prescription relates to effective metrics and which thermodynamic and geometric properties follow after calibration. A smooth matter profile coupled to the equilibrium scalar density of DRG with local running governed by the renormalization group defines a diagnostic source shape. Its Arnowitt-Deser-Misner (ADM) normalization and extremal mass are calibration conditions. The pressure consistency check determines whether the prescribed source can support the calibrated metric. A fit of the source profile to the density family $\rho(r)=\rho_0\exp[-(r/r_c)^n]$ selects $n=2$. Its associated metric best matches the direct DRG source among the tested members. Four quantitative results follow for this calibrated family:
\begin{enumerate}
\item a nonmonotonic relation between mass and temperature with a maximum Hawking temperature and an extremal endpoint at zero temperature,
\item a static lapse diagnostic whose trough value scales as $\mathcal{Z}_{\rm tr}\propto(M-\Mmin)^{-1/2}$ near the extremal mass $\Mmin$, confirmed across all tested families and the Hayward and Bardeen controls,
\item a static integral of coordinate time between the horizons, which depends on the regulator and is not an echo prediction,
\item a crossing of the area entropy at $M_{\rm cross}/M_0=0.707107$, whose limit $1/\sqrt2$ is geometric.
\end{enumerate}

The thermodynamic halt and the scaling at the lapse trough are not independent mechanisms but two late-time consequences of the same degeneration of the horizons near extremality. That observation motivates, but does not by itself establish, an mechanism for information recovery.

The discussion of quantum information is restricted to diagnostics. It provides benchmarks for the capacity inferred from area without a calculation of the density matrix of the radiation or of an island saddle. In that sense, the present analysis is complementary to recent developments based on quantum extremal surfaces (QES) and island constructions \cite{Penington2020, Almheiri2019}, emphasizing instead singularity regularization, endpoint physics and interior geometry. The resulting picture preserves a smooth, semiclassical event horizon and is compatible with the absence of firewall-like structure at the horizon scale at the level of the effective geometry considered here. Recent work emphasizes that regular static metrics need not be dynamically formed and that their inner horizons may be unstable \cite{Bueno2024Formation,Bueno2025Formation,Bonanno2025Cauchy}. For this reason the source reconstruction and collapse calculations with imposed caps below are reported as diagnostics, not as closure of formation or stability.

The remaining sections follow the sequence of the calculations. Section~\ref{sec:framework} reviews the DRG framework. Section~\ref{sec:compact} analyzes compact objects, from stars to regular black holes. Section~\ref{sec:thermo} derives the modified thermodynamics. Section~\ref{sec:barrier} discusses the static lapse diagnostic. Section~\ref{sec:echoes} evaluates the regulated integral of coordinate time between the horizons. Section~\ref{sec:implications} discusses semiquantitative benchmarks for the information capacity, including the crossing of the area entropy. Section~\ref{sec:conclusion} summarizes. Technical details are given in the appendices.

\section{The density-responsive framework}
\label{sec:framework}

DRG is a extension with a scalar field of general relativity in which geometry couples to the ambient matter density through a single scalar field. The framework was introduced in \cite{Drobczyk2025_1} as an effective description of late-time acceleration. For compact objects, the prescription for the source, metric calibration and order of differentiation and scale substitution are specified below. No cosmological fit or ultraviolet matching is established here. The application of the framework to dark matter phenomenology is developed in \cite{Drobczyk2025_2}. This section summarizes the elements needed for the analysis of regular black holes.

For the auxiliary stellar diagnostic, the physical metric is
\begin{equation}
  \tilde{g}_{\mu\nu} = \Omega^2(\rho_m)\, g_{\mu\nu},
  \label{eq:conformal}
\end{equation}
where $g_{\mu\nu}$ is an auxiliary metric. For the stellar diagnostic at fixed scale below, $\Omega(\rho_m)$ is a prescribed conformal factor that depends on density, not a universal force law derived from the scalar potential alone. Appendix~\ref{app:conformal} states this additional prescription and its restricted domain. The prescribed conformal factor is
\begin{equation}
  \ln\Omega(\rho_m)
  = A\left[
     \ln\!\left(\frac{\rho_m}{\rho_m+M_U^4}\right)
     +\frac{M_U^4}{\rho_m+M_U^4}
     \right],
  \label{eq:omega}
\end{equation}
with the historical benchmark $A=0.024$ and fixed $M_U=5.84\,\mathrm{meV}$ in the auxiliary stellar calculation. These values are diagnostic inputs, not a new cosmological determination or a microscopic matching of a black hole core. The black hole construction below separately uses a local running ansatz that depends on density and an imposed normalization of the extremal mass. Neither follows from the conformal prescription for stars at fixed scale.

At fixed $M_U$ and $\rho_m\gg M_U^4$, Equation~\eqref{eq:omega} gives $\Omega\to1$ with corrections of order $(M_U^4/\rho_m)^2$. The suppression $|\Omega-1|<4\times10^{-229}$ derived in Appendix~\ref{app:conformal} for the stellar density range pertains to that diagnostic at high density at fixed scale only. It cannot be transferred to the black hole ansatz in which $M_U^4$ is replaced by a running scale that depends on density. The conformal factor $\Omega$ is not itself the regularization mechanism. Rather, $\Omega$ encodes the density dependence of the coupling between the scalar and matter. A full demonstration that the coupled dynamics of the scalar and gravity prevent an state of infinite density requires a collapse solution, which is not yet available. The analysis uses a calibrated effective geometry with a regular core of de~Sitter type of finite curvature. A bounded scalar contribution $W$, defined in Section~\ref{subsec:drg_bh}, does not bound the total density $\rho_m+W$. Regularity of this effective geometry is an imposed property, not a consequence of that scalar bound alone.

\paragraph{Frame conventions.}
The conformal relation alone does not identify an action in the Einstein frame. The symbol $\tilde g_{\mu\nu}$ denotes the declared physical metric of each diagnostic. Dynamical equivalence of frames is not inferred without a specified action \cite{Faraoni2004}. The black hole metric is specified independently below.

\section{Density-responsive compact objects}
\label{sec:compact}

The properties of compact objects are examined through two separate calculations. For configurations supported by pressure such as stars, the relativistic equations of hydrostatic equilibrium are solved directly. A calibrated analytical metric models the proposed regular black hole state after complete collapse. The two calculations are separate diagnostics. No dynamical map from the stellar sequence to the black hole family is assumed.

\subsection{The stellar branch: modified Tolman-Oppenheimer-Volkoff equations}
\label{subsec:tov}

For static, spherically symmetric configurations supported by pressure, the fluid obeys hydrostatic balance in the physical metric $\tilde{g}_{\mu\nu}$. Fluid elements supported by pressure are not freely falling test particles. Within the approximation for the source stated in Appendix~\ref{subsec:tov_integration}, the equations take the standard Tolman-Oppenheimer-Volkoff (TOV) form \cite{Oppenheimer1939, Tolman1939}
\begin{equation}
  \frac{d\tilde{m}}{dr} = 4\pi r^2 \rho_m,
  \label{eq:tov1}
\end{equation}
\begin{equation}
  \frac{dp_m}{dr}
  = -\frac{(\rho_m + p_m)(\tilde{m} + 4\pi r^3 p_m)}
          {r(r - 2\tilde{m})}\,,
  \label{eq:tov2}
\end{equation}
with an equation of state $p_m = p_m(\rho_m)$ for closure. The adopted polytropic form is $p_m = K\,\rho_m^\Gamma$ with $K = 0.1$, $\Gamma = 2.0$ in Planck units ($G=c=\hbar=k_B=1$). This is not intended as an astrophysical model but as a theory diagnostic that demonstrates the existence of a maximum mass and the onset of gravitational instability.

\subsection{Numerical results: the relation between mass and radius}
\label{subsec:stellar_results}

\begin{figure}[t]
  \centering
  \includegraphics[width=0.7\textwidth]{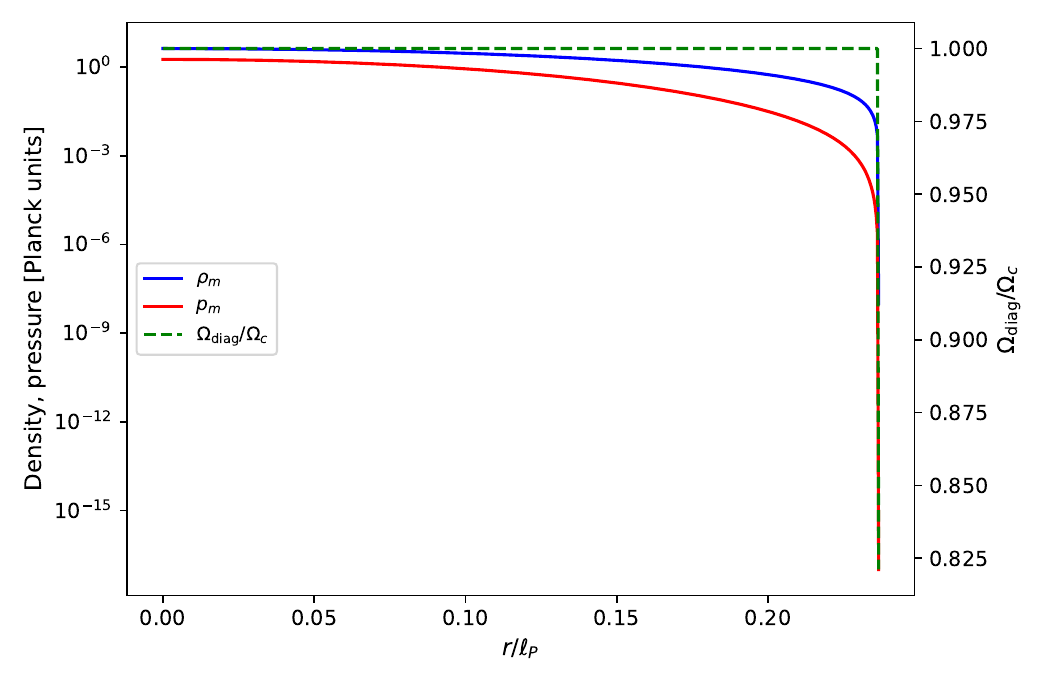}
  \caption{Interior profiles of a configuration near the maximum mass on the stellar branch, obtained from the TOV
  equations with a polytropic equation of state
  ($K = 0.1$, $\Gamma = 2.0$, Planck units).
  The matter density $\rho_m(r)$ (blue, solid) and pressure
  $p_m(r)$ (red, solid) refer to the left axis.
  The dashed green curve shows the conformal factor $\Omega(r)$
  (normalized to its central value, right axis).
  The conformal factor tracks the density profile, confirming that it is maximal at the core and decreases smoothly outward.
  The diagnostic value $M_U^4=10^{-4}\rho_P$ makes the variation
  visible. For the physical fixed infrared scale and the plotted densities
  down to $10^{-8}\rho_P$, $|\Omega-1|<4\times10^{-229}$.}
  \label{fig:star_profile}
\end{figure}

\begin{figure}[t]
  \centering
  \includegraphics[width=0.7\textwidth]{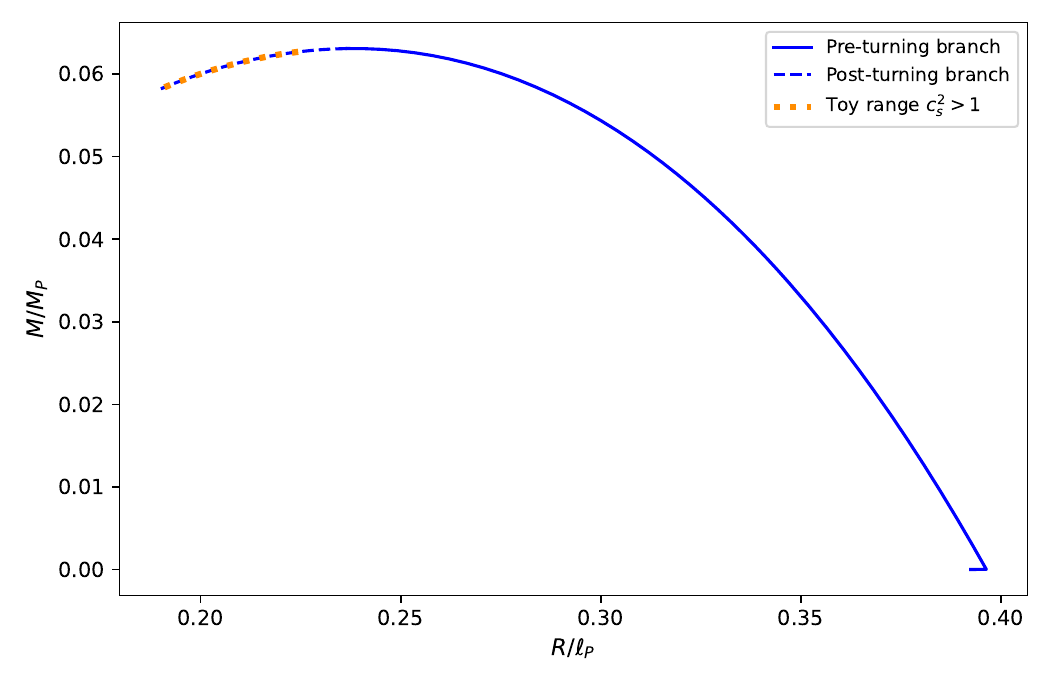}
  \caption{Relation between mass and radius for the stellar branch.
  The solid line shows the stable branch and the dashed line
  the unstable branch beyond
  $M_{\max} \approx 0.063\,\MPl$.
  The orange dotted portion has central $c_s^2>1$ and is retained
  only to display the continuation of the toy sequence.
  These configurations serve as theory diagnostics and are not intended as models of neutron stars.}
  \label{fig:mass_radius}
\end{figure}

Equations~\eqref{eq:tov1} and \eqref{eq:tov2} are integrated for central densities in the range $\rho_c \in [10^{-6},\, 10]\,\rho_P$, where $\rho_P$ is the Planck density. Figure~\ref{fig:star_profile} shows the interior density and pressure profiles for a configuration near the maximum mass. The resulting relation between mass and radius is shown in Figure~\ref{fig:mass_radius}. The sequence exhibits a well-defined turning point at
\begin{equation}
  M_{\max} \approx 0.063\,\MPl
  \quad\text{at}\quad
  \rho_c \approx 4.3\,\rho_P\,,
  \label{eq:mmax}
\end{equation}
separating a stable branch ($dM/d\rho_c > 0$) from an unstable one. The turning point marks radial instability of this toy sequence. It does not determine the endpoint of collapse.

The numerical value of $M_{\max}$ depends on the polytropic stiffness parameter $K$ and carries no direct observational significance. A realistic microphysical equation of state would require a separate stellar calculation. The turning point reported here belongs to the specified toy equation of state. Its sound speed satisfies $c_s^2=dp_m/d\rho_m=0.2\rho_m$ in Planck units, so the part of the scan with $\rho_c>5\,\rho_P$ is acausal and is marked as such.

\subsection{Construction of a DRG-motivated regular black hole}
\label{subsec:drg_bh}

A separate effective branch describes black holes with a regular core of de~Sitter type \cite{Shapiro1983, Ansoldi2008}. Its density shape is motivated by the DRG scalar sector and then calibrated. It is not obtained by evolving the stellar sequence above.

\paragraph{Source convention.}
Let $X=\rho_m$, $E=M_U^4$ and $W=AE^2/(E+X)$. The density used here follows the prescription $\rho_\Phi^{\rm eff}=W-X(\partial W/\partial X)_E$, with $p_\Phi=-W$. Differentiation is performed at fixed $E$ before inserting the local running ansatz. Varying $E(X)$ before differentiation is a different prescription. For comparison, a conserved-current fluid with $X=mn$, fixed particle mass $m$, number density $n$ and energy density $W(X,E)$ at fixed $E$ instead has $\rho_\Phi=W$ and $p_\Phi=n(\partial W/\partial n)_E-W=XW_X-W$. That is a different prescription for the source and is not substituted for the diagnostic used here. This choice defines a density diagnostic, not a complete off-shell action for matter or a static stress tensor.

\paragraph{Core asymptotics of the diagnostic.}
With dimensionless local running $M_U^4(\mu)/\rho_\star=(\mu/\mu_\star)^{4\gamma}$ and $\mu/\mu_\star\sim(\rho_m/\rho_\star)^{1/4}$, the dimensionless density is $x=\rho_m/\rho_\star$, with $\rho_\star=\rho_P$ below. The equilibrium scalar density becomes
\begin{equation}
  \frac{\rho_\Phi^{\rm eff}}{\rho_\star}
  = \frac{A x^{2\gamma}}{x+x^\gamma}
  \,\frac{x^\gamma+2x}{x^\gamma+x}\,.
  \label{eq:rho_phi_rg}
\end{equation}
Its behavior at high density is $\rho_\Phi^{\rm eff}/\rho_\star\sim2A x^{2\gamma-1}$. The constant limit occurs only at $\gamma=1/2$. The historical choice $\gamma=0.501$ instead gives slow growth $x^{0.002}$, not a strict density bound. In Appendix~\ref{subsec:calibration_logic}, the calculation matching the endpoint scales gives $\gamma\simeq0.497636$ with the unreduced Planck mass used here or $0.491798$ with the reduced Planck mass as endpoint scale. These give negative asymptotic powers, not a nonzero constant limit. The value $0.501$ is retained as a historical benchmark for the source calibration and fit. The endpoint estimates are tested separately in Appendix~\ref{subsec:calibration_logic}. No value of $\gamma$ supplies the missing static stress tensor.

\paragraph{Smooth DRG black hole.}
The adopted Gaussian matter profile $\rho_m(r) = \rho_c\exp(-r^2/\sigma^2)$ determines the total density $\rho_{\rm tot}(r) = \rho_m(r) + \rho_\Phi^{\rm eff}(\rho_m(r))$. The width $\sigma$ is calibrated by requiring that the integrated ADM mass $M = 4\pi\int_0^\infty r^2\rho_{\rm tot}\,dr$ match the target value (ADM error $< 10^{-14}$). The central matter density $\rho_c = 0.156\,\rho_P$ is calibrated so that the extremal mass matches the imposed benchmark
\begin{equation}
  \Mmin = \sqrt{1 + A/2}\,\MPl \simeq 1.006\,\MPl\,.
  \label{eq:mmin_theory}
\end{equation}
The physical line element is
\[
d\tilde s^2=-f(r)\,dt^2+f(r)^{-1}\,dr^2+r^2d\Omega_2^2,
\]
where $r$ is the areal radius, $d\Omega_2^2$ is the unit-sphere metric and $t$ is normalized at infinity. The units satisfy $G=c=\hbar=k_B=1$, with $\MPl=G^{-1/2}$ the unreduced Planck mass. The diagnostic metric function then follows algebraically,
\begin{equation}
  f(r) = 1 - \frac{2\,m(r)}{r}\,,\qquad
  m(r) = 4\pi\int_0^r \tilde r^2\,\rho_{\rm tot}(\tilde r)\,d\tilde r\,.
  \label{eq:f_smooth}
\end{equation}
This construction uses no TOV equation and encounters no horizon singularity, but it is not a solution of the full static DRG field equations. The resulting geometry has two horizons for all $M > \Mmin$ and a bounded central Kretschmann scalar $K(0) \simeq 45.7\,\MPl^4$. It recovers the Schwarzschild limit at large mass.

\paragraph{Pressure consistency.}
At the calibrated center, the density prescription gives $\rho_{\rm tot}=0.16466$ and the algebraic DRG pressure is $p_{\rm DRG}=-0.00678$ in Planck units. However, the metric with a single function in Equation~\eqref{eq:f_smooth} requires $p_r=-\rho_{\rm tot}=-0.16466$. The relative mismatch is $|p_{\rm DRG}-p_r|/\rho_{\rm tot}=0.959$. The smooth profile is used only to motivate and rank density shapes.

\paragraph{Fit of the source profile to the metric family.}
A fit of the smooth DRG profile with consistent ADM mass against the parametric density family $\rho(r) = \rho_0\exp[-(r/r_c)^n]$ identifies $n = 2$ as the effective metric best matching the direct DRG source at all tested mass ratios ($M/\Mmin = 1.05, 1.20, 2.00, 10.00$). The $n=2$ family is then calibrated by fixing $\rho_0$ such that the extremal mass $M_{\rm ext}$ of the family equals $\Mmin$. The mass function for this family is
\begin{equation}
  m(r) = M\,P\!\left(\tfrac{3}{n},\,(r/r_c)^n\right),
  \qquad
  r_c = \left[\frac{Mn}{4\pi\rho_0\,\Gamma(3/n)}\right]^{1/3},
  \label{eq:drg_family}
\end{equation}
where $P(a,z)=\Gamma(a)^{-1}\int_0^z t^{a-1}e^{-t}\,dt$ is the regularized lower incomplete gamma function and $\Gamma(a)$ is the complete gamma function. Here the first argument $3/n$ of $P$ is not a cosmological scale factor. The index $n$ specifies the shape of the density profile and $r_c$ is its radial scale that depends on mass. The Gaussian matter profile in the smooth construction is an ansatz, not a solution of the DRG field equations. The fit of the source profile to the metric family does not prove that DRG dynamically produces the $n=2$ family. What the fit does establish is that, of the tested parametric completions, $n=2$ best matches the calibrated density shape. The subsequent thermodynamic and causal results depend on the calibrated $n=2$ family, not on the Gaussian ansatz itself.

\paragraph{Validation by reconstruction of the source.}
To test whether the calibrated $n=2$ family has the source structure expected of a regular black hole, its effective stress tensor is reconstructed from Einstein's equations in the Schwarzschild gauge with one metric function. In this gauge the relation $p_r = -\rho$ holds identically for any metric of the form $f(r) = 1 - 2m(r)/r$, so the condition $\rho + p_r = 0$ at the horizons is a structural property of the gauge rather than a dynamical prediction (Appendix~\ref{subsec:app_inverse_source}). The additional information comes from the reconstructed tangential pressure. The reconstructed source becomes precisely isotropic as $r\to0$, since $|p_t-p_r|/\rho=(n/2)(r/r_c)^n\to0$. Together with the identity $p_r=-\rho$, this gives a full de~Sitter core at the center. These properties confirm the regular de~Sitter form of the effective metric, but do not repair the mismatch with the independently prescribed algebraic DRG pressure above. Technical details are given in Appendix~\ref{subsec:app_inverse_source}.

\paragraph{Calibrated metric and scope of the results.}
All subsequent thermodynamic, lapse and calculations of coordinate time are performed on this calibrated $n=2$ effective metric family motivated by DRG. The smooth profile serves as a diagnostic bridge between the DRG density dependence and the calibrated regular family. It is not a direct existence proof. The Hayward-type control~\cite{Hayward2005} and the Bardeen metric~\cite{Bardeen1968} are retained as literature benchmarks, not as the primary model.

The specific contribution of the density prescription is the comparison of source profiles and its test of pressure consistency, followed by quantitative diagnostics of the selected, calibrated metric. The exponent at a double root and the capacity ratio inferred from area at large mass are not new consequences of DRG. Neither the source fit nor the extremal calibration establishes an advantage in dynamical formation, stability or information recovery.

Table~\ref{tab:metric_comparison} compares the calibrated $n=2$ family with Hayward and Bardeen. The qualitative endpoint structure is shared, but its coefficients depend on the metric family. The exponent at the lapse trough is $-0.500$ for all families within the numerical fit accuracy. The regulated integral of coordinate time is also compared below, but is not promoted to an echo observable.

\paragraph{Exploratory check of boundedness.}
An exploratory Eulerian calculation with finite volumes probes the density prescription under an imposed bounce factor. It is not an autonomous DRG collapse. The gravity source, a Newtonian force with the pressure correction of the TOV equation, is explicitly multiplied by the imposed bounce factor $(1-\rho_{\rm tot}/\rho_{\rm crit})$ with $\rho_{\rm crit}=(1+A/2)\,\rho_P$. Clipping and numerical viscosity are part of the scheme. For a compact Gaussian initial profile on a fixed grid ($N=400$, $\sigma_{\rm IC}=1$), nine central densities from $\rho_c = 0.02$ to $0.20\,\rho_P$ were evolved. The two lowest-density cases ($\rho_c = 0.02$ and $0.03$) remain subcritical and never develop cells with $1-2m/r<0$. The remaining seven runs with $\rho_c \ge 0.04$ evolve to configurations with bounded density with such cells persisting to $t=100$ within this capped scheme. In the run with $\rho_c=0.20\,\rho_P$ documented in Appendix~\ref{subsec:app_eulerfv}, such cells are already present in the initial data. The runs neither remove the imposed clipping nor establish regularity of an uncapped evolution. They show numerical boundedness of the capped model only and do not establish formation of the calibrated static geometry.

\begin{table}[t]
  \centering
  \caption{Comparison of calibrated metrics of regular black holes.
  The DRG $n=2$ family is the primary calibrated effective model,
  selected by the fit of the source profile to the metric family.
  The Hayward-type and Bardeen metrics serve as controls. The lapse column gives the exponent $\alpha$ in $\mathcal{Z}_{\rm tr}\propto(M-\Mmin)^{\alpha}$.
  All models are fixed to
  $\Mmin = \sqrt{1+A/2}\,\MPl$.
  The last column reports the regulated integral over the static cavity at
  $M=10\,\Mmin$ and $\epsilon=10^{-6}$ and is not an echo observable.}
  \label{tab:metric_comparison}
  \begin{tabular}{lccccc}
    \toprule
Model & Type & $\Mmin/\MPl$ & $\Tmax/\MPl$ & lapse $\alpha$ & $\Delta t_{\rm cav}/M$ \\
    \midrule
    DRG $n=2$ & fitted to source & 1.006 & 0.0255 & $-0.500$ & $52.6$ \\
    DRG $n=3$ & family scan & 1.006 & 0.0278 & $-0.500$ & $52.8$ \\
    DRG $n=4$ & family scan & 1.006 & 0.0295 & $-0.500$ & $53.0$ \\
    Hayward & literature & 1.006 & 0.0252 & $-0.500$ & n/a \\
    Bardeen & literature & 1.006 & 0.0243 & $-0.500$ & n/a \\
    \bottomrule
  \end{tabular}
\end{table}

\section{Modified black hole thermodynamics}
\label{sec:thermo}

The calibrated metric defined in Section~\ref{subsec:drg_bh} determines the thermodynamic properties analyzed below. Singularity regularization leads to a qualitative modification of the Hawking temperature, replacing the divergent endpoint of Schwarzschild evaporation with a extremal endpoint at zero temperature.

\subsection{The relation between mass and temperature}
\label{subsec:TH}

\begin{figure}[t]
  \centering
  \includegraphics[width=0.7\textwidth]{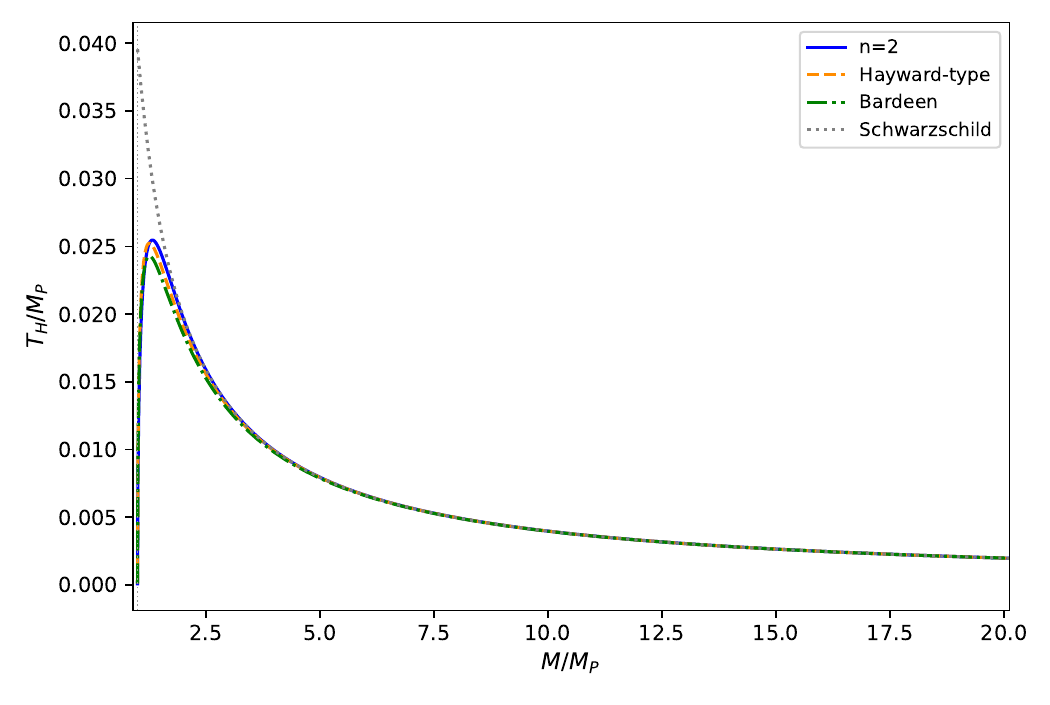}
  \caption{Hawking temperature as a function of mass for the
  calibrated DRG $n=2$ family (solid blue), Hayward-type control (dashed orange) and Bardeen control (dash-dotted green),
  compared with Schwarzschild (dotted gray).
  All regular models are calibrated to the
  same extremal mass $\Mmin \approx 1.006\,\MPl$.
  The common endpoint does not imply quantitative equality of the
  temperature curves.
  The vertical dotted line marks $\Mmin$.}
  \label{fig:thermo}
\end{figure}

The Hawking temperature is determined by the surface gravity at the outer horizon $r_h$
\begin{equation}
  T_H = \frac{1}{4\pi}\,f'(r_h)\,.
  \label{eq:TH_def}
\end{equation}
For the calibrated $n=2$ effective metric family motivated by DRG (Equation~\eqref{eq:drg_family}), the horizon equation $f(r_h)=0$ and its derivative $f'(r_h)$ are evaluated numerically. For the Hayward-type control used here, with fixed core length $\ell$ and $f(r)=1-2Mr^2/(r^3+\ell^3)$, explicit parametric relations are available. This is a Hayward-type family with a fixed denominator scale. The original form with denominator $r^3+2ML^2$ instead has a fixed length $L$ of the central curvature. The relation $\ell^3=2ML^2$ means that those two choices define different mass sequences. Here $\ell$ is held fixed throughout.
\begin{equation}
  M(r_h) = \frac{r_h^3 + \ell^3}{2r_h^2}\,,
  \label{eq:M_of_rh}
\end{equation}
\begin{equation}
  T_H(r_h) = \frac{1}{4\pi}\,
  \frac{r_h^3 - 2\ell^3}{r_h(r_h^3 + \ell^3)}\,.
  \label{eq:TH_of_rh}
\end{equation}
The resulting relation between mass and temperature, shown in Figure~\ref{fig:thermo}, differs qualitatively from the Schwarzschild result $T_H = 1/(8\pi M)$. Instead of diverging at small mass, the temperature reaches a maximum
\begin{equation}
  \Tmax = 0.02547\,\MPl
  \quad\text{at}\quad M = 1.332\,\MPl
  \label{eq:TH_max}
\end{equation}
and then decreases, reaching zero at the extremal mass $\Mmin$. For masses below the peak, the black hole enters a cooling phase in which further mass loss leads to a decrease in temperature. This thermodynamic turnaround prevents a runaway endpoint in the quasistatic evaporation equation. Dynamical stability is a separate question. The nonmonotonic behavior is a generic consequence of singularity regularization \cite{Frolov2016}.

\subsection{Recovery of the Schwarzschild limit}
\label{subsec:gr_limit}

For the Hayward-type control with $M \gg \MPl$, a Taylor expansion in the small parameter $(\ell/r_h)^3 \ll 1$ gives
\begin{equation}
  T_H(M) = \frac{1}{8\pi M}
  \left[1 - \frac{\ell^3}{4 M^3}
  + \mathcal{O}\!\left(\frac{\ell^6}{M^6}\right)\right].
  \label{eq:TH_GR}
\end{equation}
The leading term is the standard Schwarzschild temperature. The correction is suppressed by $(\ell/M)^3 \sim (\MPl/M)^3$, which for an astrophysical black hole of mass $M \sim 10\,M_\odot$ gives a relative deviation of order $10^{-118}$.

\begin{figure}[t]
  \centering
  \includegraphics[width=0.7\textwidth]{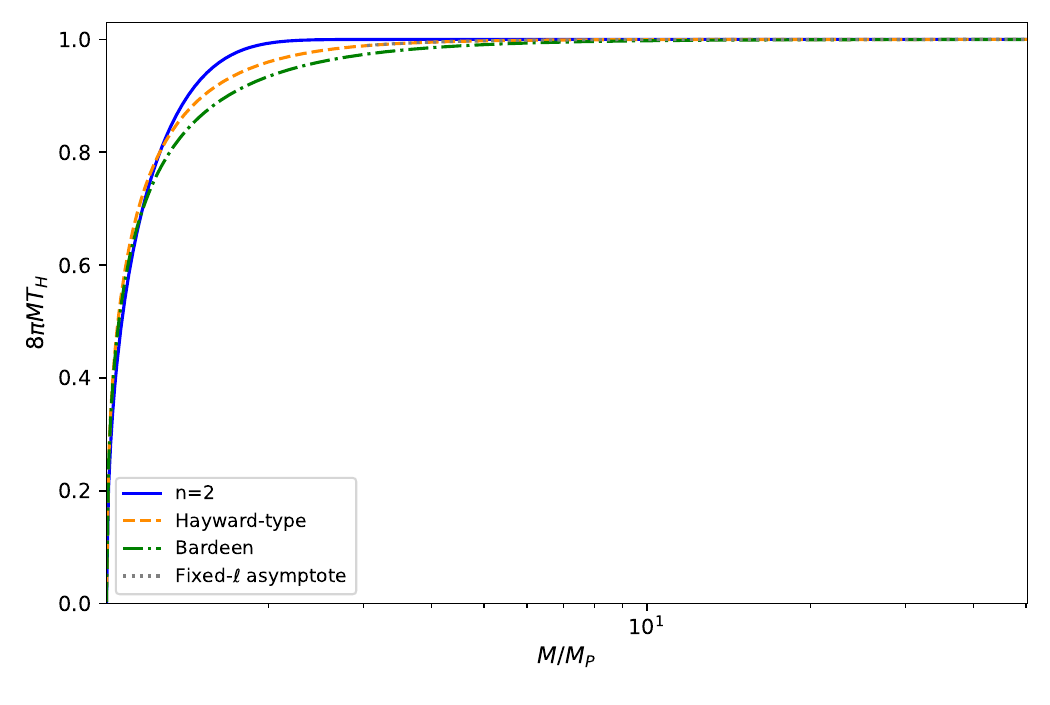}
  \caption{The diagnostic ratio $8\pi M\,T_H(M)$ for the
  DRG $n=2$ family and the Hayward and Bardeen controls.
  All approach unity for large masses. The dotted gray asymptote,
  shown for $M\geq3\,\Mmin$, is Equation~\eqref{eq:TH_GR} for the Hayward-type control only. The $n=2$ corrections are exponential,
  whereas the Bardeen correction starts at relative order $M^{-2}$.}
  \label{fig:gr_diagnostic}
\end{figure}

The diagnostic ratio $8\pi M\,T_H(M)$ is shown in Figure~\ref{fig:gr_diagnostic} for the DRG $n=2$ family and the Hayward and Bardeen controls. All three approach unity at large mass, confirming consistency with general relativity. The temperature values remain family dependent. At $M=1.01\,\Mmin$, the Hayward-type and Bardeen controls exceed the $n=2$ temperature by $21.1\%$ and $19.4\%$, respectively. At $M=1.5\,\Mmin$ they lie $3.6\%$ and $7.0\%$ below it. The common results are the qualitative turnover and the asymptotic Schwarzschild limit, not percent-level universality.

\subsection{Area capacity and the temperature integral}
\label{subsec:entropy_correction}

A temperature correction need not generate a logarithmic entropy term. In the example with quantized area of Bagchi, Ghosh and Sen~\cite{Bagchi2024Landauer}, a temperature correction of relative order $M^{-2}$ generates a logarithm in $\int dM/T$. Equation~\eqref{eq:TH_GR} instead contains a relative $M^{-3}$ correction at fixed $\ell$. The formal integral of $1/T_H$ is
\begin{equation}
 S_{\rm int}(M)\equiv\int^M\frac{d\bar M}{T_H(\bar M)}
 =4\pi M^2-\frac{2\pi\ell^3}{M}
 +\mathcal O(\ell^6/M^4)+S_{\rm c}.
 \label{eq:entropy_integral_expansion}
\end{equation}
Here $S_{\rm c}$ is an arbitrary additive constant. The exact result for the Hayward-type control is $S_{\rm int}(r_h)=\pi r_h^2-2\pi\ell^3/r_h+S_{\rm c}$. This formal integral contains no logarithmic term. This statement concerns the specified classical family of metrics, not quantum corrections to an unspecified underlying theory.

For the primary $n=2$ family at fixed $\rho_0$, the corrections at large mass are exponentially suppressed, with $x_h=(r_h/r_c)^2\propto M^{4/3}$ and $1-P(3/2,x_h)\sim2\sqrt{x_h}e^{-x_h}/\sqrt{\pi}$. Thus this family also supplies no logarithm in its temperature integral at large mass. Its integral is $S_{\rm int}=4\pi M^2+\mathrm{const}$ plus exponentially suppressed terms. In both cases the coefficient of a logarithm inferred from this classical calculation is zero. A quantum logarithmic coefficient is not calculated.

The formal integral is not automatically the physical entropy. With $S_{\rm area}=\pi r_h^2$, differentiation of $r_h=2m(r_h,M)$ gives the exact identity for the metric family
\begin{equation}
 T_H\,dS_{\rm area}=\mathcal C_h\,dM,\qquad
 \mathcal C_h=\left.\frac{\partial m(r,M)}{\partial M}\right|_{r=r_h},
 \label{eq:area_first_law_factor}
\end{equation}
where other family parameters are held fixed. For the fixed-$\ell$ control, $\mathcal C_h=r_h^3/(r_h^3+\ell^3)$. For the $n=2$ family at fixed $\rho_0$, $\mathcal C_h=P(3/2,x_h)-(2/3)x_h^{3/2}e^{-x_h}/\Gamma(3/2)$. The dependence of the source on mass enters the first law, as discussed by Ma and Zhao~\cite{MaZhao2014FirstLaw}. $S_{\rm area}$ is retained only as the geometric capacity used below. Determining a physical entropy and its quantum corrections requires a specified action and a consistent treatment of source variations. The physical entropy cannot be obtained by identifying $S_{\rm int}$ with $S_{\rm area}$.

\subsection{Evaporation dynamics and the halt of mass loss}
\label{subsec:evaporation}

\begin{figure}[t]
  \centering
  \includegraphics[width=0.98\textwidth]{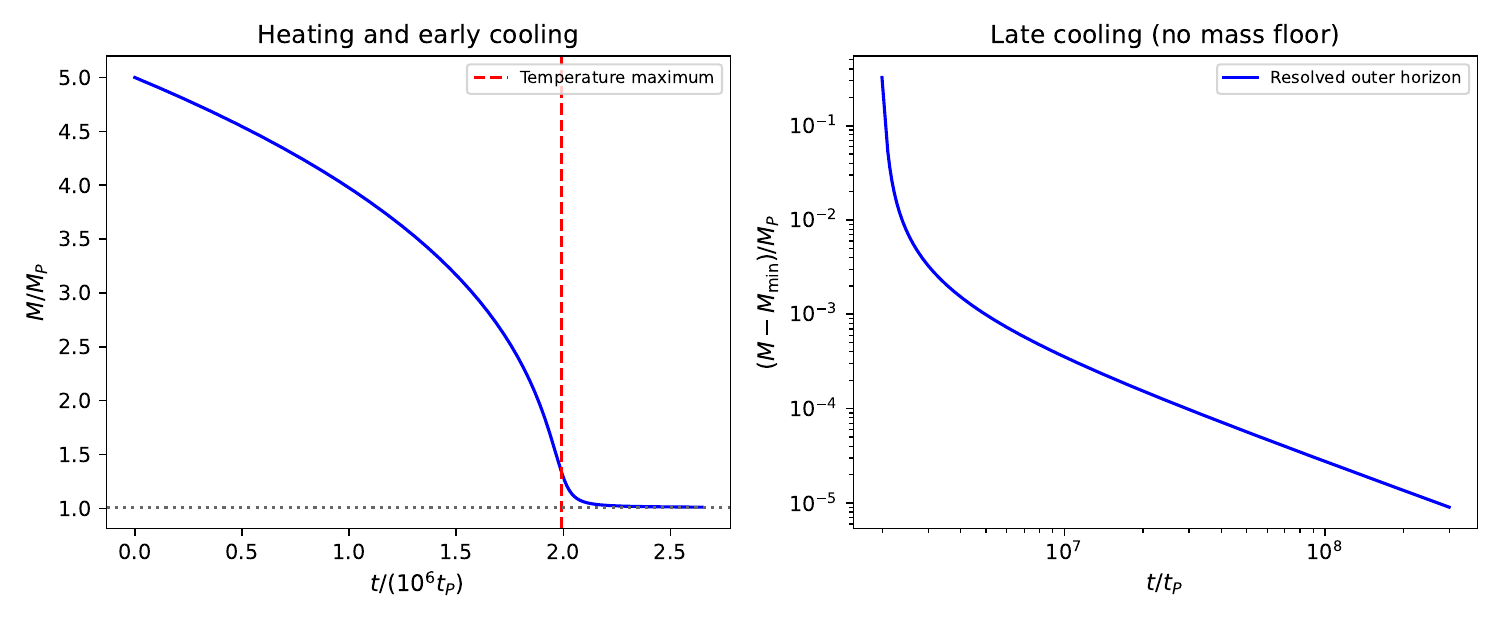}
  \caption{Mass as a function of time for the evaporation of a
  regular black hole with initial mass $M_0 = 5\,\MPl$.
  Left: heating and early cooling. The vertical dashed line marks
  the temperature maximum at $M=1.332\,\MPl$. The horizontal dotted line marks $\Mmin$. Right: the long cooling tail, plotted
  as the remaining mass excess on logarithmic axes. The outer horizon is resolved at each evaluation and no mass floor is used.
  The endpoint is approached asymptotically.}
  \label{fig:evaporation}
\end{figure}

The rate of mass loss for a nonrotating, uncharged black hole is \cite{Page1976}
\begin{equation}
  \frac{dM}{dt} = -\mathcal{G}\,\sigma\,A_H\,T_H^4\,,
  \label{eq:dMdt}
\end{equation}
where $A_H = 4\pi r_h^2$ is the horizon area and $\mathcal{G}$ denotes frequency-averaged transmission effects. For the numerical integration shown in Figure~\ref{fig:evaporation}, the blackbody coefficient for two polarizations is $\sigma=\pi^2/60$ and the transmission factor is set to $\mathcal G=1$. This fixes only the illustrative time normalization. Realistic greybody factors require a perturbation calculation \cite{Rincon2020Greybody}. The scaling of Equation~\eqref{eq:scaling} is independent of this prefactor. For a Schwarzschild black hole, decreasing mass leads to increasing temperature, resulting in a runaway process that terminates in an explosive final burst. The nonmonotonic $T_H(M)$ of the regular black hole changes this late-time behavior.

\begin{figure}[t]
  \centering
  \includegraphics[width=0.7\textwidth]{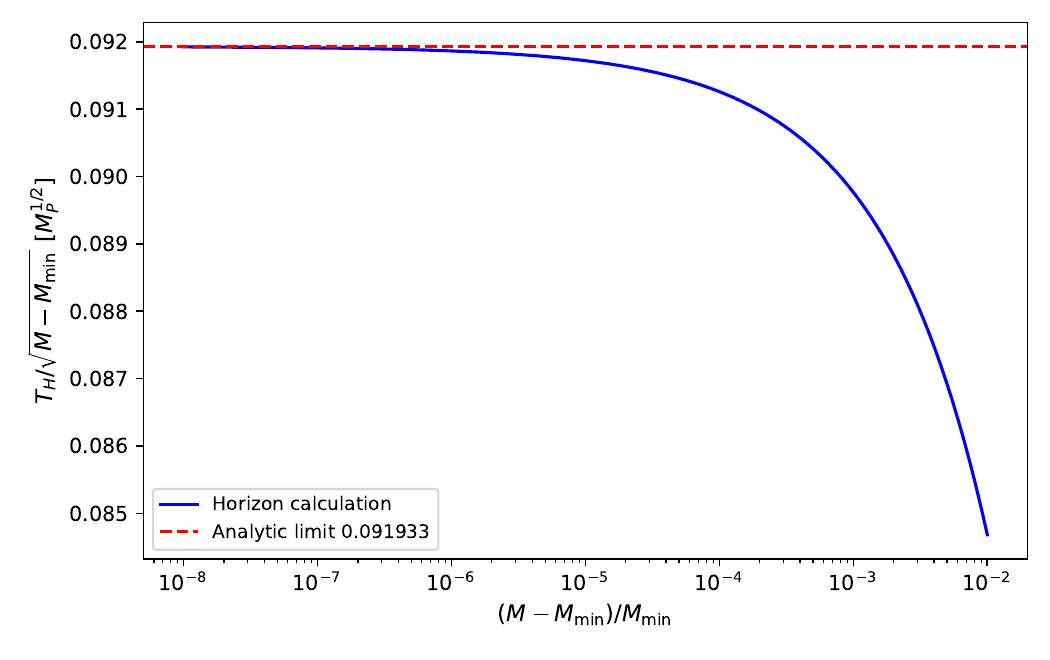}
  \caption{Scaling near extremality of the Hawking temperature.
  The ratio $T_H/\sqrt{M-\Mmin}$ approaches the analytic
  coefficient $0.0919330\,\MPl^{1/2}$ as the mass excess
  decreases. Its drift at finite mass explains the smaller
  coefficient obtained from a fit over a wider window.}
  \label{fig:scaling}
\end{figure}

Near extremality, the temperature follows the universal scaling law
\begin{equation}
  T_H \simeq C\,\sqrt{M - \Mmin}\,,
  \label{eq:scaling}
\end{equation}
where $C=0.0919330\,\MPl^{1/2}$ for the calibrated $n=2$ family (Figure~\ref{fig:scaling}). This is the asymptotic coefficient, $C=\sqrt{-2f_M f_{rr}}/(4\pi)$, evaluated at the double root. Here $f_M=\partial f/\partial M$ and $f_{rr}=\partial^2 f/\partial r^2$, with the central density held fixed in $f_M$. A fit over a finite interval over $(M-\Mmin)/\Mmin=10^{-6}$ to $10^{-2}$ instead gives approximately $0.087\,\MPl^{1/2}$ and must not be identified with that limiting coefficient. Inserting the limiting square root law for $T_H$ into Equation~\eqref{eq:dMdt} gives $dM/dt \propto -(M - \Mmin)^2$, which has the solution
\begin{equation}
  M(t) = \Mmin + \frac{1}{c_{\rm ev}\,t + \mathrm{const}}
  \label{eq:M_of_t}
\end{equation}
for a positive constant $c_{\rm ev}$. The mass approaches $\Mmin$ only asymptotically as $t \to \infty$. Evaporation never reaches $M < \Mmin$ in finite time.

Figure~\ref{fig:evaporation} shows the numerical integration of Equation~\eqref{eq:dMdt} for an initial mass $M_0 = 5\,\MPl$. The solution exhibits two phases: an initial heating phase (Phase~I), during which the temperature increases as $M$ decreases, followed by a cooling phase (Phase~II) beyond the peak at $M = 1.332\,\MPl$, where the temperature and rate of mass loss decrease together. The solution approaches a extremal endpoint at zero temperature.

\subsection{Properties of the extremal endpoint}
\label{subsec:remnant}

The quasistatic endpoint is an extremal configuration that does not radiate at the Planck scale. Stability and dynamical formation are not inferred from $T_H=0$. The calibrated properties of the endpoint are summarized in Table~\ref{tab:remnant}.

\begin{table}[t]
  \caption{Properties of the calibrated extremal endpoint.
  All quantities are in Planck units.}
  \label{tab:remnant}
  \begin{ruledtabular}
  \begin{tabular}{lcc}
    Property & Symbol & Value \\
    \hline
    Endpoint mass     & $\Mmin$     & $1.006\,\MPl$ \\
    Extremal radius  & $r_\star$   & $1.597\,\lPl$ \\
    Temperature      & $T_{\rm rem}$ & $0$ (exactly) \\
    Horizon area     & $A_{\rm rem}$ & $32.0\,\lPl^2$ \\
    Bekenstein-Hawking entropy & $\Srem$ & $8.01$ nats \\
  \end{tabular}
  \end{ruledtabular}
\end{table}

If the area law is applied literally, $\Srem \approx 8.0$ corresponds to an scale set by the area entropy of roughly 12~bits. For a progenitor of $10\,M_\odot$ the initial area entropy is $S_i = 4\pi(M_i/\MPl)^2 \simeq 1.05\times10^{79}$, so that $\Srem / S_i \simeq 7.6\times10^{-79}$ (Section~\ref{subsec:species}). This is not a microscopic state count. The resulting restriction on the interpretation of capacity is discussed in Section~\ref{sec:implications}.

\section{The lapse diagnostic of the static interior}
\label{sec:barrier}

The modified thermodynamics of Section~\ref{sec:thermo} describes the external evolution of the black hole. The static lapse between the inner and outer horizons describes a different aspect of the geometry. The quantity below is a coordinate-dependent diagnostic of the geometry with two horizons, not by itself an observable redshift.

\subsection{Origin: the geometry with two horizons}
\label{subsec:barrier_origin}

Singularity regularization produces a core of de~Sitter type with $f(0)>0$. For a member with a negative interior minimum and asymptotic flatness $f(r)\to1$ as $r\to\infty$, continuity then requires at least two zeros of $f(r)$: an inner (Cauchy) horizon $r_-$ and an outer (event) horizon $r_+$. Between these horizons $f(r) < 0$. The lapse diagnostic is defined by
\begin{equation}
  \mathcal{Z}(r) = \frac{1}{\sqrt{|f(r)|}}\,,
  \label{eq:Z}
\end{equation}
which diverges at both coordinate horizons.

Geometries with two horizons also arise in the Kerr and Reissner-Nordstr\"om families, where the inner horizon is generated by angular momentum or charge. In those cases the structure with two horizons vanishes in the nonrotating, uncharged limit. By contrast, the inner horizon of a regular black hole is a direct consequence of the regularized center ($f(0)>0$) and persists for all values of $M > \Mmin$, independent of spin or charge. The associated extremal mass $\Mmin$ is fixed by the calibration, not by a conserved quantity.

This diagnostic is computed from the physical metric of the black hole, declared separately, through $f(r)$ alone. It does not use the conformal estimate at fixed scale from the stellar calculation.

\begin{figure}[t]
  \centering
  \includegraphics[width=0.7\textwidth]{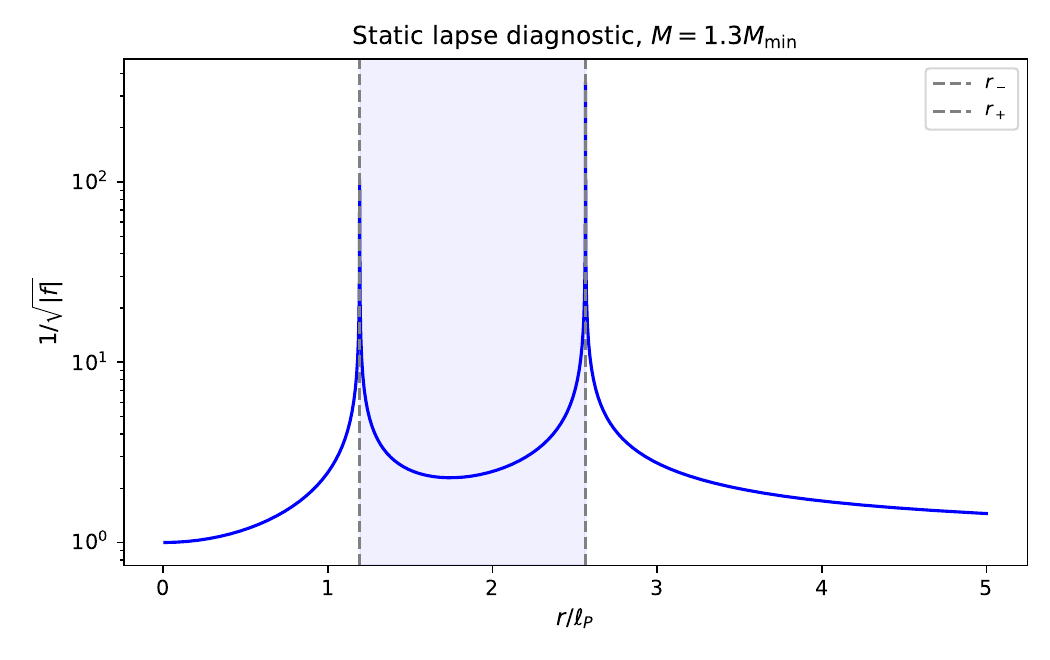}
  \caption{The static lapse quantity $\mathcal{Z}(r) = 1/\sqrt{|f(r)|}$
  for a regular black hole with $M = 1.3\,\Mmin$.
  The quantity diverges at both the inner horizon $r_-$ and
  the outer horizon $r_+$ (dashed gray lines), with an interior
  trough between them (shaded). The finite plotted peaks are
  samples near the poles, not finite physical maxima.}
  \label{fig:barrier}
\end{figure}

Figure~\ref{fig:barrier} shows $\mathcal{Z}(r)$ for a nonextremal configuration with $M = 1.3\,\Mmin$. The coordinate divergences at $r_\pm$ are visible.

\subsection{Scaling near the extremal limit}
\label{subsec:barrier_scaling}

\begin{figure}[t]
  \centering
  \includegraphics[width=0.7\textwidth]{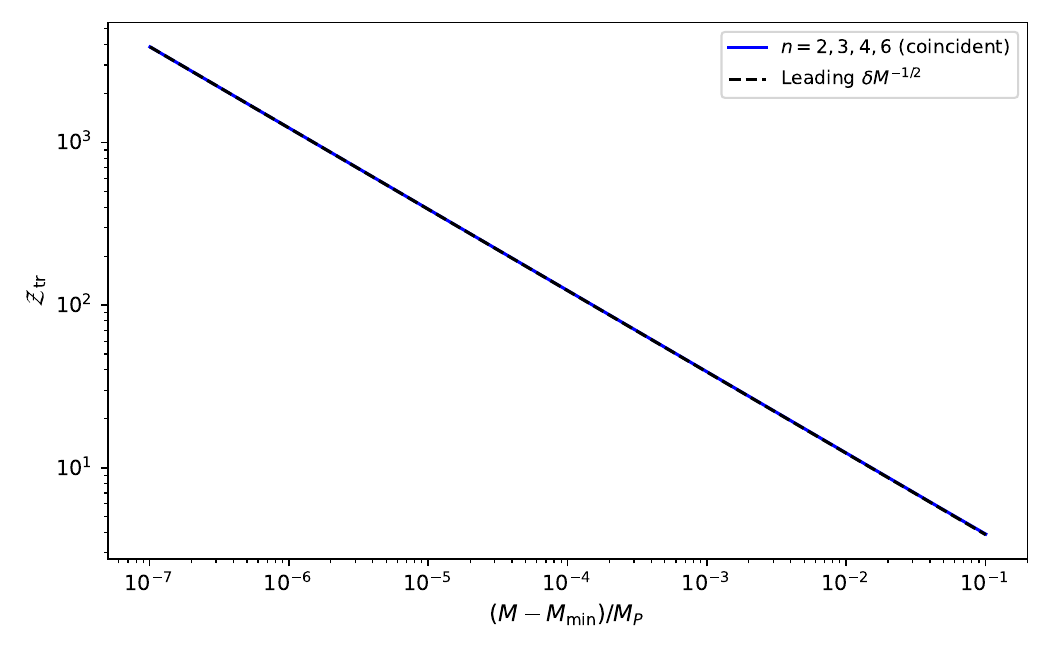}
  \caption{Value at the lapse trough $\mathcal{Z}_{\rm tr}$ versus
  $\delta M=M-\Mmin$ for the calibrated $n=2,3,4,6$ families.
  The four curves coincide because their mass scaling gives $f(r_{\rm tr})=1-(M/\Mmin)^{2/3}$.
  The dashed line shows the generic scaling at a double root
  $\mathcal{Z}_{\rm tr}\propto\delta M^{-1/2}$.
  This is a trough diagnostic without a cutoff on distance from a horizon.}
  \label{fig:barrier_scaling}
\end{figure}

As the black hole evaporates and $M \to \Mmin$, the two horizons approach each other. Let $r_{\rm tr}$ minimize $f(r)$ between the horizons. Then $-f(r_{\rm tr})=\max|f(r)|$ is positive and proportional to $M-\Mmin$ near a nondegenerate double root. Thus the trough value $\mathcal Z_{\rm tr}=\min\mathcal Z(r)$, not a finite maximum at either horizon, scales as
\begin{equation}
  \mathcal{Z}_{\rm tr}
  \propto \frac{1}{\sqrt{M - \Mmin}}\,.
  \label{eq:Z_scaling}
\end{equation}
For the normalized families with exponential density, the exact relation $f(r_{\rm tr})=1-(M/\Mmin)^{2/3}$ gives the common curve in Figure~\ref{fig:barrier_scaling} and its limiting exponent $-1/2$. The same limiting exponent follows from the nondegenerate double roots of the Hayward-type and Bardeen controls. This universality follows from the generic structure of the double root of $f(r)$ at extremality and does not depend on the details of the interpolation. This particular diagnostic is evaluated at the trough and uses no cutoff on distance from a horizon. It must be distinguished from a redshift sampled at a fixed proper distance from a horizon and from the integral of coordinate time of Section~\ref{sec:echoes}, which is regulated by a fixed fraction of the horizon gap.

The divergence of $\mathcal{Z}$ is a statement about the static Schwarzschild-like chart. It does not by itself determine signal propagation, information transfer or dynamics across the horizons.

\subsection{Synchronization with the thermodynamic halt}
\label{subsec:synergy}

The scaling of the lapse trough and the vanishing of the Hawking temperature are not independent phenomena. Both originate from the same geometric property: the merging of the inner and outer horizons at the extremal configuration, defined by $f(r_\star) = 0$ and $f'(r_\star) = 0$ simultaneously.

A Taylor expansion of $f(r)$ around $r_\star$ shows that $f'(r_h) \propto \sqrt{M - \Mmin}$ and $-f(r_{\rm tr}) \propto (M - \Mmin)$. The temperature $T_H = f'(r_h)/(4\pi) \propto \sqrt{M - \Mmin}$ vanishes at the same rate at which the lapse diagnostic diverges, $\mathcal{Z}_{\rm tr} \propto 1/\sqrt{M - \Mmin}$. The two mechanisms cannot be detuned. They are mathematically coupled through the structure of $f(r)$ near its degenerate root.

This synchronization ensures a smooth quasistatic approach to the extremal endpoint within the calibrated model. Neither a final burst of radiation nor an abrupt disappearance of a horizon occurs.

\paragraph{Stability of the Cauchy horizon.}
The inner (Cauchy) horizon of a regular black hole with two horizons is generically subject to the instability due to mass inflation discovered by Poisson and Israel~\cite{PoissonIsrael1990} and further analyzed by Ori~\cite{Ori1991}. Perturbations crossing the inner horizon experience exponential blueshifting, which can drive the local mass function and curvature invariants to large values even in initially regular geometries \cite{CarballoRubio2022_inner}. This instability has direct implications for the results of this work. If mass inflation destroys the smooth geometry at the inner horizon, the static diagnostics between the horizons need not persist in a dynamical, perturbed spacetime. Recent work reinforces the point that regularity of a background metric does not ensure stability of the Cauchy horizon \cite{Bonanno2025Cauchy,Frolov2026MassInflation,Liu2026MassInflation}. In the limit near extremality $M \to \Mmin$ studied here, the inner and outer areal radii merge. This is not a statement about an invariant spatial volume. The static time slices are not spacelike between the horizons. No suppression of mass inflation is established by this limit. The dynamical behavior of the inner horizon requires a perturbative or numerical treatment of the regular interior beyond this framework based on an effective metric. The kinematic curve relating mass and temperature depends only on the calibrated branch of the outer horizon, although dynamical instability could alter the physical endpoint.

\section{Regulated coordinate time between the horizons}
\label{sec:echoes}

An integral of coordinate time between the horizons can be tempting to interpret as a delay of a echo of a gravitational wave. That identification is not justified by the static metric alone. An echo prediction requires a spin-2 perturbation potential and boundary conditions. This calculation is retained only as a diagnostic of the static chart that depends on the regulator. For a generic regular metric with mass function $m(r)$ the integral is
\begin{equation}
  \Delta t_{\mathrm{cav}}
  = 2 \int_{r_-}^{r_+} \frac{dr}{|f(r)|}\,.
  \label{eq:echo_integral}
\end{equation}
Because $f(r) \to 0$ at both horizons, the integral diverges logarithmically at both ends and the result depends on the integration limits. A cutoff $\epsilon$ is introduced as a fraction of the horizon gap, with integration from $r_-+\epsilon(r_+-r_-)$ to $r_+-\epsilon(r_+-r_-)$. The cutoff is a coordinate regulator, not a reflecting surface.

Equation~\eqref{eq:echo_integral} is evaluated numerically for $\epsilon = 10^{-3}$ to $10^{-8}$ (see Appendix~\ref{subsec:app_echo} for details). Figure~\ref{fig:echo_delay} shows the result. The shaded band spans the full cutoff range. At $M=10\,\Mmin$ and $\epsilon=10^{-6}$ the calibrated $n=2$ model gives $\Delta t_{\rm cav}/M=52.6$. This quantity defined in the static chart is not converted into a detector echo time.

\begin{figure}[t]
  \centering
  \includegraphics[width=0.7\textwidth]{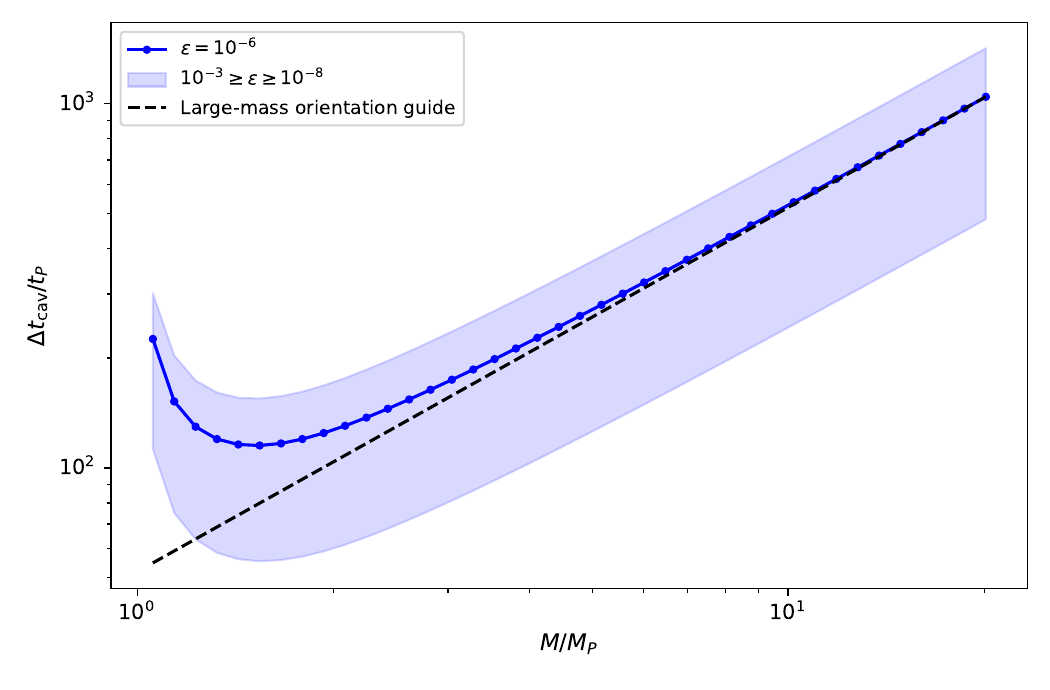}
  \caption{Static diagnostic based on coordinate time between the horizons as a
  function of black hole mass.
  Circles show the numerical evaluation of the cavity integral
  in Equation~\eqref{eq:echo_integral} at the reference
  cutoff $\epsilon = 10^{-6}$.
  The shaded band spans $\epsilon = 10^{-3}$ to $10^{-8}$.
  The dashed line is an orientation guide $\Delta t\propto M/\MPl$ for large mass, normalized at the largest plotted mass. It is neither a global fit nor an echo formula.}
  \label{fig:echo_delay}
\end{figure}

The calculation supplies neither a perturbative transfer function nor an observational prediction. Studies of regular black hole observables illustrate the additional input required for that step \cite{Urmanov2025Observational}.

\section{Benchmarks for information capacity}
\label{sec:implications}

The area of the outer horizon provides an scale for information capacity, but the static geometry does not determine a radiation density matrix. The purpose of this section is not to claim a resolution of the information paradox, but to clarify what these results do and do not establish. In particular, a rigorous treatment would require a calculation in quantum field theory of the entanglement structure of Hawking radiation in the regularized geometry, which lies beyond the scope of this work.

\subsection{Single geometric origin and the crossing of the area entropy}
\label{subsec:page}

The thermodynamic halt of Section~\ref{sec:thermo} and the divergent lapse diagnostic of Section~\ref{sec:barrier} should not be viewed as two independent mechanisms. They are two late-time consequences of the same degeneration of the horizons near extremality. The vanishing of $T_H$ and the growth of $\mathcal{Z}_{\rm tr}$ are locked by the same structure of the double root of $f(r)$. This establishes a common kinematic origin, not information recovery.

The information paradox is quantified through the von Neumann entropy of the emitted radiation, $S_R = -\mathrm{Tr}(\rho_R \ln \rho_R)$, which measures the entanglement between the radiation and the remaining black hole \cite{Page1993, Page2013}. In Hawking's original calculation, each emitted quantum is a member of an entangled pair, with one particle escaping as radiation and its partner falling behind the horizon. As a result, the radiation entropy $S_R$ increases monotonically \cite{Hawking1976}. When the black hole disappears, the internal degrees of freedom vanish, leaving the radiation in a mixed state. This is a manifest violation of unitarity.

If the evolution is unitary, the global state remains pure and the entanglement entropies of complementary subsystems agree. In a semiclassical treatment, the Bekenstein-Hawking entropy $S_{\rm BH} = A_H/4$ \cite{Bekenstein1973} provides the relevant benchmark for information capacity for the description of the black hole exterior. Accordingly, one expects the radiation entropy to turn over once it becomes comparable to the decreasing black hole entropy and to return toward zero as the evaporation approaches its final state. This trajectory, which first rises and then falls, is the Page curve. The central challenge for any resolution of the information paradox is to identify a physical mechanism that produces the turnover.

\paragraph{Capacity inferred from area.}
Let $S_{\rm area}(M)=\pi r_h^2(M)$ and define the accumulated area loss $S_{\rm loss}(M)=S_{\rm area}(M_0)-S_{\rm area}(M)$. These two geometric curves cross when
\begin{equation}
  S_{\rm loss}(M_{\rm cross})=S_{\rm area}(M_{\rm cross}).
  \label{eq:S_hawk}
\end{equation}
This is not an island formula. Neither curve is identified with the von Neumann entropy of Hawking radiation.

Table~\ref{tab:page_curve} shows the results for a range of initial masses on the calibrated DRG $n=2$ family. The crossing ratio
\begin{equation}
  \frac{M_{\rm cross}}{M_0} = 0.707107 \simeq \frac{1}{\sqrt2}
  \label{eq:page_ratio}
\end{equation}
approaches $1/\sqrt2$ at large mass because $S_{\rm area}\propto M^2$. The small differences between families, $0.708671$ and $0.713555$ for the Hayward-type and Bardeen controls at $M_0=5\,\MPl$, are also geometric and are not a prediction of unitary evolution.

\begin{table}[t]
  \centering
  \caption{Crossing of the area entropy for the DRG $n=2$ family.
  The ratio $M_{\rm cross}/M_0$ agrees with $1/\sqrt2$ to better than $10^{-7}$ for all listed initial masses. The table uses $S_0\equiv S_{\rm area}(M_0)$.
  The remnant entropy fraction $\Srem/S_0$ decreases rapidly
  with $M_0$.}
  \label{tab:page_curve}
  \begin{tabular}{lccc}
    \toprule
    $M_0/\MPl$ & $S_{\rm area}(M_0)$ & $M_{\rm cross}/M_0$ & $\Srem/S_0$ \\
    \midrule
    5 & 314 & 0.707 & 2.55\% \\
    10 & 1257 & 0.707 & 0.64\% \\
    20 & 5027 & 0.707 & 0.16\% \\
    50 & 31\,416 & 0.707 & 0.025\% \\
    \bottomrule
  \end{tabular}
\end{table}

For large $M_0$ the crossing ratio approaches $1/\sqrt{2} \simeq 0.707$, the value set by the crossing of $S_{\rm loss} = S_{\rm area}$ for any metric with $S_{\rm area} \propto M^2$. The approximate universality of $M_{\rm cross}/M_0$ is a geometric consequence of the relation between area and entropy, not a prediction specific to DRG. The calibrated geometry adds a well-defined extremal endpoint. Evaporation approaches $\Mmin$ asymptotically, with $\Srem/S_0 < 2.6\%$ even for $M_0 = 5\,\MPl$ and $< 0.03\%$ for $M_0 = 50\,\MPl$. The small residual area does not determine how quantum information is stored or exported.

Appendix~\ref{subsec:app_qes_saddles} evaluates a functional inspired by QES as an exploratory sensitivity study. Its bulk term and candidate surfaces are imposed rather than derived, so it does not validate an island saddle.

\begin{figure}[t]
  \centering
  \includegraphics[width=0.7\textwidth]{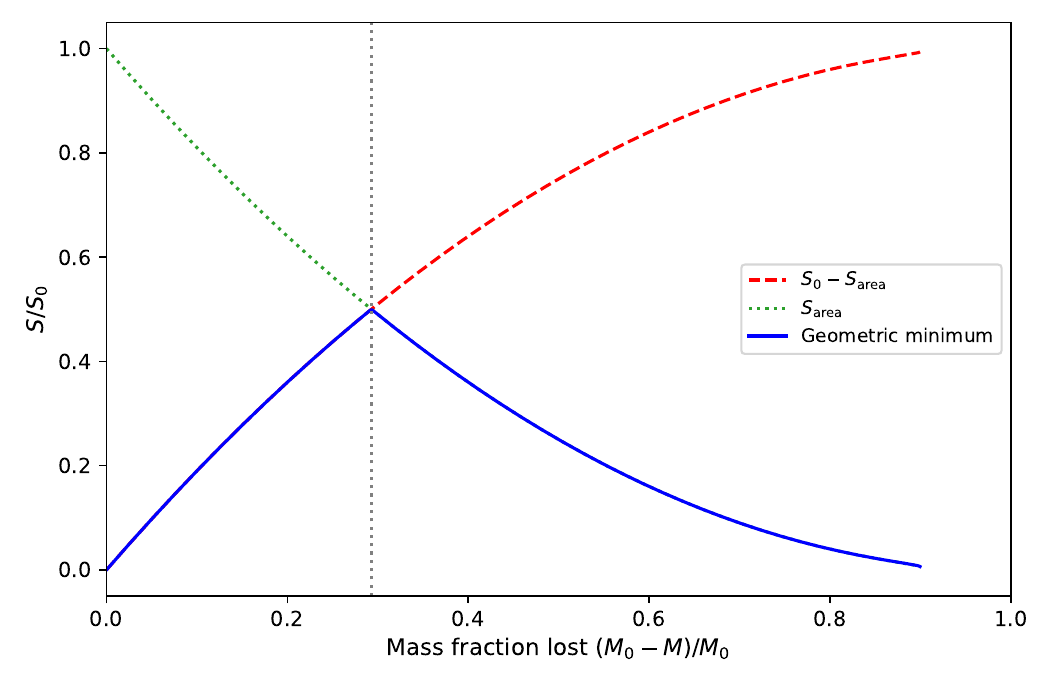}
  \caption{Crossing of the area entropy for the calibrated regular black hole of the DRG $n=2$ family with $M_0 = 10\,\MPl$.
  The horizontal axis shows the mass fraction lost,
  $(M_0 - M)/M_0$, so that time runs from left to right.
  The dashed red line is the accumulated area loss and the dotted green line the remaining area entropy. The blue line is their
  minimum and serves only as a visual crossing benchmark. The
  crossing occurs at $M_{\rm cross}/M_0=0.707107$. All quantities
  are normalized to $S_{\rm area}(M_0)$.}
  \label{fig:page_curve}
\end{figure}

Figure~\ref{fig:page_curve} shows the crossing of the capacities inferred from area for $M_0 = 10\,\MPl$. The accumulated area loss increases while the remaining area entropy decreases. Their minimum is plotted only to display the crossing.

No statement about a Page time, radiation entropy or information flux follows from this construction.

\subsection{Why no transfer bound follows from the lapse diagnostic}
\label{subsec:transfer}

Combining the coordinate quantity $\mathcal Z$ with the quasistatic evaporation law does not yield a coordinate-invariant information rate and presupposes a microscopic channel that has not been derived. For this reason no transfer bound is stated. The calculation neither proves nor excludes late-time information transfer.

\subsection{Remnant capacity and the species problem}
\label{subsec:species}

A longstanding objection to remnant scenarios is the species problem. If remnants must store all the information of arbitrarily large progenitors, their Hilbert space must be infinite-dimensional, potentially leading to pathological rates of pair production \cite{Giddings1992, Giddings1993, Susskind1995, Chen2015}.

The calibrated endpoint does not by itself resolve this objection. A stellar-mass progenitor with $M_i \sim 10\,M_\odot \simeq9.14\times10^{38}\,\MPl$ carries initial entropy $S_i = 4\pi(M_i/\MPl)^2 \simeq 1.05\times 10^{79}$ nats. The remnant entropy is $\Srem \approx 8.0$ nats (Table~\ref{tab:remnant}), so the export requirement is
\begin{equation}
  \Delta S_{\rm exp} = S_i - \Srem \simeq S_i\,,
  \label{eq:entropy_export}
\end{equation}
with
\begin{equation}
  \frac{\Srem}{S_i} \simeq 7.64\times 10^{-79},
  \qquad
  f_{\rm rad}
  \equiv \frac{\Delta S_{\rm exp}}{S_i}
  = 1 - \frac{\Srem}{S_i}
  \simeq 1\,.
  \label{eq:frad}
\end{equation}
These ratios only show that the semiclassical endpoint area is tiny relative to the progenitor area. Interpreting this as a microscopic capacity or as information export requires a quantum theory of the endpoint. No transfer rate follows from the static lapse diagnostic.

\subsection{Compatibility with a smooth horizon}
\label{subsec:firewall}

The firewall hypothesis \cite{Almheiri2013, AMPS2013} posits that unitarity requires a high-energy barrier at the event horizon, violating the equivalence principle. The argument rests on a tension between three assumptions: unitarity of evaporation, validity of semiclassical physics outside the horizon and smoothness of the horizon for an infalling observer. To purify late-time radiation, an outgoing Hawking mode must be entangled with the early radiation. However, it must also be entangled with its partner mode immediately inside the horizon. The monogamy of entanglement forbids this tripartite structure \cite{Coffman2000}. The firewall is the proposed resolution that breaks the interior entanglement.

The calibrated metric is smooth at the outer horizon, but no radiation state or entanglement calculation is provided. Consequently the construction neither resolves nor materially changes the firewall argument. It supplies a regular background on which that quantum problem could be posed.

\section{Conclusion}
\label{sec:conclusion}

This work examined the thermodynamics and static interior geometry of regular black holes motivated by the scalar sector of DRG. The equilibrium scalar density with local running was used to construct a diagnostic source profile. A fit of this profile to a family of effective metrics selected the $n=2$ member, with extremal mass calibrated to $\Mmin \simeq 1.006\,\MPl$. This construction provided the basis for four quantitative results.

The Hawking temperature reaches a maximum $\Tmax = 0.02547\,\MPl$ at $M=1.332\,\MPl$ and falls to zero at the extremal endpoint $\Mmin$. Both the maximum and the approach to the endpoint according to a square root law occur in all tested families. Their coefficients depend on the metric. The maximum temperatures for $n=3$ and $n=4$ exceed the $n=2$ value by $9.3\%$ and $15.7\%$, respectively.

At the lapse trough, the static diagnostic scales as $\mathcal{Z}_{\rm tr} \propto (M - \Mmin)^{-1/2}$. The exponent follows analytically for the $n=2$ family and all tested controls. Its divergence and the vanishing temperature have the same geometric origin: the double root formed when the two horizons merge at extremality.

The regulated integral between the horizons yields $\Delta t_{\rm cav}/M=52.6$ at $M=10\,\Mmin$ and $\epsilon=10^{-6}$. This value characterizes the static cavity and does not predict an echo delay.

At $M_{\rm cross}/M_0=0.707107\simeq1/\sqrt2$, half the initial area entropy remains. Together with the small fraction at the endpoint, this provides a geometric benchmark for capacity. These geometric quantities do not establish a Page curve, an island saddle or unitary recovery of information. The exploratory functional inspired by QES does not supply the missing quantum state.

Reconstruction of the source from the effective metric recovers an isotropic central core. The directly prescribed source nevertheless retains a relative pressure mismatch of $0.959$ and does not solve the full static DRG equations. The collapse runs using finite volumes and an imposed bounce factor remain bounded, but rely on clipping and numerical viscosity. They do not establish formation of the calibrated geometry.

Connecting these static results to a physical endpoint requires a complete source action and a treatment of collapse in numerical relativity. Such a calculation would test formation and stability and would be needed to compute gravitational waveforms. Whether the endpoint at zero temperature is dynamically realized as a stable remnant remains open. Determining its information content also requires the entanglement structure of Hawking radiation in the regularized background. A full QES calculation with dynamical bulk entropy would test whether a Page curve arises and connect these geometric diagnostics to the quantum information problem.

\section*{Data availability}

The data that support the findings of this article are not publicly available. The data are available from the author upon reasonable request.

\appendix

\section{Auxiliary conformal prescription at fixed scale}
\label{app:conformal}

This appendix records the auxiliary prescription used for the stellar diagnostic. It is not a first-principles derivation of a universal matter force. In particular, a conformal relation does not by itself identify the gravitational or matter action.

\subsection{Prescribed differential relation}
\label{subsec:force_matching}

Conservation of the total stress-energy constrains the sum of matter and scalar stresses but does not determine their exchange vector without a source action and a matter dictionary. The relation at fixed scale is adopted rather than derived from conservation
\begin{equation}
  \frac{d\ln\Omega}{d\rho_m}
  = \frac{A\,M_U^8}{\rho_m\,(\rho_m + M_U^4)^2}\,.
  \label{eq:app_master}
\end{equation}
This relation is used only on the positive-density interval of the stellar diagnostic. No extension to a generic source supported by pressure or to $M_U(\rho_m)$ is implied.

\subsection{Integration and normalization}
\label{subsec:integration_norm}

Using the partial-fraction decomposition
\[
\frac{1}{\rho_m(\rho_m + M_U^4)^2} = \frac{1}{M_U^8}\!\left( \frac{1}{\rho_m} - \frac{1}{\rho_m + M_U^4} - \frac{M_U^4}{(\rho_m + M_U^4)^2} \right),
\]
direct integration of Equation~\eqref{eq:app_master} yields the general solution
\begin{equation}
  \ln\Omega(\rho_m)
  = A\left[
     \ln\!\left(\frac{\rho_m}{\rho_m + M_U^4}\right)
     + \frac{M_U^4}{\rho_m + M_U^4}
     \right] + \mathcal{C}\,.
  \label{eq:app_omega_general}
\end{equation}
The chosen normalization is $\Omega\to1$ as $\rho_m\to\infty$, giving $\mathcal C=0$. For $A>0$ this expression instead tends to zero as $\rho_m\to0$. It does not provide a smooth match to a finite constant conformal factor of the vacuum. Its use is restricted to the stellar integration interval with positive density. A global conformal completion of that diagnostic is not claimed.

\paragraph{Parameter values.}
Here $A=0.024$ and $M_U=5.84\,\mathrm{meV}$ are historical diagnostic inputs. The scale is held fixed in this auxiliary conformal calculation. The locally running source used in the black hole comparison is a separate prescription. Neither calculation imports a cosmological fit as validation of the construction of compact objects.

\paragraph{Limit at high density.}
For densities relevant to compact objects ($\rho_m \gg M_U^4$), expanding in $x \equiv M_U^4/\rho_m \ll 1$ gives
\begin{equation}
  \ln\Omega = -\frac{A}{2}\,x^2
  + \mathcal{O}(x^3)\,.
\end{equation}
At the Planck density, $x \simeq 5\times10^{-122}$ with the unreduced Planck mass. Over the range $10^{-8}\,\rho_P$ to $10\,\rho_P$ this gives $|\Omega - 1| < 4\times10^{-229}$. This estimate applies only at fixed $M_U$ in the stellar diagnostic. It neither establishes the black hole source with running scale nor derives the regular metric.

\section{Numerical and analytical methods}
\label{app:methods}

\subsection{Units, frames and conventions}
\label{subsec:units}

The calculations use Planck units ($G = c = \hbar = k_B = 1$) and metric signature $(-,+,+,+)$. Einstein's equations read $G_{\mu\nu} = 8\pi\,T_{\mu\nu}$.

For the black hole branch, the declared physical metric uses the form with a single function
\begin{equation}
  d\tilde{s}^2
  = -f(r)\,dt^2 + f(r)^{-1}\,dr^2 + r^2\,d\Omega_2^2\,,
\end{equation}
where $r$ is the areal radius and the mass function $m(r)$ is defined by $f(r) = 1 - 2m(r)/r$. The total gravitational mass is $M = \lim_{r\to\infty} m(r)$. The Hawking temperature is
\begin{equation}
  T_H = \frac{\kappa}{2\pi}\,,
  \qquad
  \kappa = \frac{1}{2}\,f'(r_h)\,,
\end{equation}
where $r_h$ is the radius of the outer event horizon.

The stellar branch instead uses $d\tilde s^2=-e^{2\nu(r)}dt^2+[1-2\tilde m(r)/r]^{-1}dr^2 +r^2d\Omega_2^2$, with an independent lapse $\nu(r)$. The TOV and thermodynamic calculations use their declared physical metrics. No field equations in the Einstein frame or dynamical frame equivalence are inferred from Equation~\eqref{eq:conformal}. The matter equation of state is a polytrope $p_m = K\,\rho_m^\Gamma$ with $K = 0.1$ and $\Gamma = 2.0$ (in Planck units). This choice is not intended to model realistic matter in neutron stars but to produce a stellar sequence with a well-defined maximum mass at the Planckian scales relevant here.

\subsection{Source prescription and calibration}
\label{subsec:calibration_logic}

The logic of the black hole construction used in the main text is as follows.

\paragraph{Density diagnostic motivated by DRG.}
The scalar density follows from the chosen algebraic prescription, Equation~\eqref{eq:rho_phi_rg}. The independently imposed benchmark scale is $\Mmin^{\rm bench}=\sqrt{1+A/2}\,\MPl$. The Gaussian radial profile, its width and its central density are additional choices. The match of the ADM mass and the match of the extremal mass are imposed as calibration conditions. The pressure consistency check in Section~\ref{subsec:drg_bh} shows why this is not a direct existence proof.

\paragraph{Fit of the source profile to the metric family.}
A comparison with the parametric family $\rho(r) = \rho_0\exp[-(r/r_c)^n]$ identifies the $n=2$ member as the effective metric that matches best. This family is then calibrated by fixing $\rho_0$ such that $M_{\rm ext} = \Mmin^{\rm bench}$. All subsequent thermodynamic, lapse, coordinate time and area calculations are performed on this calibrated $n=2$ family. The Hayward-type and Bardeen metrics are retained as controls, not as the primary model.

The robustness test used throughout the paper is family stability at fixed $\Mmin^{\rm bench}$. Results that hold in the $n=2,3,4,6$ scan and the Hayward and Bardeen comparison are interpreted as shared across the tested families rather than as artifacts of a particular interpolation.

\paragraph{Sensitivity to the running exponent.}
An explicit estimate using the endpoint scales follows from the power-law ansatz $M_U(\mu)=\mu_{\rm UV}(\mu/\mu_{\rm UV})^\gamma$. Matching $M_U(\mu_{\rm IR})=5.84\times10^{-3}\,\mathrm{eV}$ at $\mu_{\rm IR}=1.44\times10^{-33}\,\mathrm{eV}$ gives
\[
\gamma=\frac{\ln[M_U(\mu_{\rm IR})/\mu_{\rm UV}]} {\ln(\mu_{\rm IR}/\mu_{\rm UV})}.
\]
The endpoint choices $\mu_{\rm UV}=1.22\times10^{28}\,\mathrm{eV}$ and $2.435\times10^{27}\,\mathrm{eV}$ correspond to unreduced and reduced Planck masses, respectively. This is a conditional matching estimate, not a derivation of a microscopic running law. The calibration of the source and comparison of the root mean square (RMS) deviation of the metric were repeated at four masses with $\gamma=0.4976357062$ and $0.4917981504$, keeping the imposed $\Mmin^{\rm bench}$ fixed. The recalibrated central matter densities are $0.1562265$ and $0.1567166$ in units of $\rho_P$. The $n=2$ member keeps the smallest metric RMS at all four tested masses in each case. Its metric RMS ranges are $0.0100$ to $0.0214$ and $0.0105$ to $0.0225$, respectively, compared with $0.00972$ to $0.02081$ for the historical benchmark. The central pressure mismatches remain $0.9586$ and $0.9581$. Neither variation repairs the failure of source consistency. The second comparison changes the exponent only and is not a conversion of all quantities to reduced Planck units. The source calibration and fit shown in this work retain $\gamma=0.501$. Once the $n=2$ effective metric and its extremal normalization are fixed, its temperature curve is independent of this source exponent and is unchanged by these comparisons.

Three components remain outside the calibrated sector: the true geometry of dynamical collapse, the perturbation problem and the quantum state that would define a physical information entropy. In these three places the analysis remains conservative.

\subsection{Stellar branch: TOV integration}
\label{subsec:tov_integration}

The TOV system (Equations~\eqref{eq:tov1} and \eqref{eq:tov2}) is solved in the physical (matter) frame using $\rho_{\rm eff} \approx \rho_m$ and $p_{\rm eff} \approx p_m$ as source terms. At the fixed scale $E=M_U^4$, the chosen prescription gives $\rho_\Phi^{\rm eff}/\rho_m\simeq2A(E/\rho_m)^2$ and $|p_\Phi|/\rho_m\simeq A(E/\rho_m)^2$ at high density. These small source ratios, rather than the smallness of $\Omega-1$ alone, justify the approximation for the source for this stellar diagnostic. They do not apply to the black hole diagnostic with running scale.

For a given central density $\rho_c$, integration starts at $r_0 = 10^{-6}$ with $\tilde{m}(r_0) = \frac{4\pi}{3}\,r_0^3\,\rho_c$ and $\rho_m(r_0) = \rho_c$. Integration terminates when $\rho_m$ drops to $\rho_{\rm surf} = 10^{-8}\,\rho_P$. The radius $R$ at this point defines the stellar radius. The total gravitational mass is $M = \tilde{m}(R)$.

The integrator is SciPy's \texttt{solve\_ivp} (\texttt{DOP853} method), with relative tolerance $2\times10^{-10}$ and absolute tolerance $10^{-12}$. A terminal event function locates the stellar surface accurately. The curve relating mass and radius is constructed by scanning $\rho_c \in [10^{-6},\,10]\,\rho_P$, extending well beyond the turning point at $\rho_c \approx 4.3\,\rho_P$.

\subsection{Black hole branch: the DRG \texorpdfstring{$n=2$}{n=2} family}
\label{subsec:app_bh}

The primary model is the effective metric family motivated by DRG with density profile $\rho(r) = \rho_0\exp[-(r/r_c)^n]$ and mass function given by Equation~\eqref{eq:drg_family}. The $n=2$ member is selected by the fit of the source profile to the metric family of Section~\ref{subsec:drg_bh}. A regular core generically violates the strong energy condition near $r = 0$, a standard feature of models of regular black holes \cite{Ansoldi2008}.

For this family, the horizon radii and Hawking temperature are computed numerically from $f(r_h) = 0$ and $T_H = f'(r_h)/(4\pi)$. The calibration parameter $\rho_0$ is fixed by matching the extremal mass to $\Mmin^{\rm bench} = \sqrt{1 + A/2}\,\MPl \simeq 1.006\,\MPl$.

For the Hayward-type control, the metric $f(r) = 1 - 2Mr^2/(r^3 + \ell^3)$ admits explicit parametric relations. The extremal radius and minimum mass are $r_\star = 2^{1/3}\ell$ and $\Mmin = 3\ell/(2^{5/3})$, giving $\ell \approx 1.065\,\lPl$ after calibration.

\subsection{Bardeen control and calibration}
\label{subsec:app_bardeen}

Validation uses the Bardeen metric $f(r) = 1 - 2Mr^2/(r^2 + g^2)^{3/2}$ with core scale $g$. Extremality occurs at $r_\star^2 = 2g^2$, giving $\Mmin = 3\sqrt{3}\,g/4$ and the calibration $g \approx 0.774\,\lPl$.

\subsection{Horizon structure}
\label{subsec:horizon_structure}

For the effective metric family motivated by DRG and the Hayward and Bardeen controls, the horizon structure follows the same pattern. An analysis of $f(r_h) = 0$ yields three cases:
\begin{itemize}
  \item $M > \Mmin$: two distinct positive roots (nonextremal regular black hole with inner and outer horizons).
  \item $M = \Mmin$: a single degenerate root (extremal black hole).
  \item $M < \Mmin$: no real positive roots. The spacetime is horizonless, with a regular core of de~Sitter type.
\end{itemize}
This confirms that the thermodynamic endpoint of evaporation ($\Mmin$) coincides with a fundamental change in causal structure. The event horizon ceases to exist for masses below this value.

\subsection{Evaporation integration}
\label{subsec:app_evap}

The integration variable is $x=\ln[(M_0-\Mmin)/(M-\Mmin)]$. The equation $dt/dx=(M-\Mmin)/|\dot M|$ is integrated with DOP853, relative tolerance $2\times10^{-10}$, from $x=0$ to $13$. At each evaluation, the two horizons are bracketed separately around the metric minimum. The selected outer root must satisfy $|f(r_h)|<2\times10^{-12}$. A failed root search terminates the calculation rather than retaining a previous radius. The temperature uses the analytic radial derivative. Independent quadratures of $dt/dx$ check the integrated times at three points.

\subsection{Integration of static coordinate time}
\label{subsec:app_echo}

The cavity integral in Equation~\eqref{eq:echo_integral} is evaluated using adaptive Gauss-Kronrod quadrature, implemented with SciPy's \texttt{quad} routine. The integration boundaries are offset by a fraction of the horizon gap, $r_{\rm lo}=r_-+\epsilon(r_+-r_-)$ and $r_{\rm hi}=r_+-\epsilon(r_+-r_-)$. Integration uses \texttt{limit}=300 and relative tolerance $2\times10^{-8}$, with the logarithm of the distance from each horizon as the integration variable.

The integral diverges logarithmically at both horizons. To quantify the resulting dependence on the regulator, the integral is evaluated for three values of $\epsilon$ spanning $10^{-3}$ to $10^{-8}$. Across the cutoff range the ratio $\Delta t_{\max}/\Delta t_{\min}$ lies between approximately $2.68$ and $2.95$ for the displayed masses. This spread quantifies regulator dependence of the static diagnostic based on coordinate time. The shaded band in Figure~\ref{fig:echo_delay} shows this range. The reference value throughout the main text is $\epsilon = 10^{-6}$.

\section{Construction of the DRG black hole}
\label{app:drg_bh}

\subsection{Smooth DRG black hole}
\label{subsec:app_smooth}

The smooth DRG black hole of Section~\ref{subsec:drg_bh} is constructed as follows. The matter profile is taken as a Gaussian, $\rho_m(r) = \rho_c\exp(-r^2/\sigma^2)$. At each radius, the DRG scalar density is computed from Equation~\eqref{eq:rho_phi_rg}. The total density $\rho_{\rm tot}(r) = \rho_m(r) + \rho_\Phi^{\rm eff}(\rho_m(r))$ is integrated to give the mass function $m(r)$ and hence $f(r) = 1 - 2m(r)/r$. At fixed $\rho_c$, setting $u=r/\sigma$ gives $M_{\rm ADM}=\sigma^3\mathcal I(\rho_c)$, where $\mathcal I=4\pi\int_0^\infty u^2\rho_{\rm tot}(u)\,du$. The shape integral is evaluated numerically to determine $\sigma=[M/\mathcal I(\rho_c)]^{1/3}$. A separate radial integration checks the ADM normalization to better than $10^{-14}$. Brent's method is used for the calibration of the central density and extremal mass, not for this width formula.

The central matter density $\rho_c = 0.156\,\rho_P$ is calibrated so that the smallest mass admitting two horizons matches $\Mmin^{\rm bench}$. At the extremal configuration, the smooth geometry has a bounded Kretschmann scalar $K(0) \simeq 45.7\,\MPl^4$ and a total central density $\rho_{\rm tot}(0) \simeq 0.165\,\rho_P$, of which the scalar contribution is $\rho_\Phi^{\rm eff}(0) \simeq 0.009\,\rho_P$. The algebraic DRG pressure, prescribed independently, has the relative central mismatch $0.959$ discussed in the main text.

\subsection{Fit of the source profile to the metric family}
\label{subsec:app_fit}

The smooth DRG density profile is compared against the calibrated parametric family $\rho(r) = \rho_0\exp[-(r/r_c)^n]$ on the radial grid consistent with the ADM normalization. The fit is performed at four mass ratios $M/\Mmin = 1.05, 1.20, 2.00, 10.00$ and ranked using diagnostics of logarithmic density and the metric function. For the $n=2$ family the metric RMS lies in the range $0.0097$ to $0.0208$, while the $n=3$ and $n=4$ families give $0.051$ to $0.125$ and $0.073$ to $0.190$, respectively.

The $n=2$ family is selected as the best effective completion.  The RMS of the logarithmic density is used as a ranking diagnostic, not as a percentage residual.

\subsection{Source consistency of the calibrated \texorpdfstring{$n=2$}{n=2} family}
\label{subsec:app_inverse_source}

For a static, spherically symmetric line element in the Schwarzschild gauge with one metric function,
\begin{equation}
  ds^2 = -f(r)\,dt^2 + \frac{dr^2}{f(r)} + r^2 d\Omega_2^2\,,
\end{equation}
Einstein's equations give the effective source components
\begin{align}
  8\pi \rho &= \frac{1 - f - r f'}{r^2}\,,
  \\
  8\pi p_r &= \frac{f - 1 + r f'}{r^2}\,,
  \\
  8\pi p_t &= \frac{r f'' + 2 f'}{2r}\,.
\end{align}
The first two expressions imply $p_r = -\rho$ identically for the inverse source of any metric of this form. For the calibrated DRG $n=2$ family this gives $\rho + p_r = 0$ exactly at both horizons. A numerical check at $M/M_{\rm ext} = 1.05,\,1.20,\,2.00$ confirms this. Using $p_t=-\rho-r\rho'/2$, the anisotropy of this family is
\begin{equation}
  \frac{|p_t-p_r|}{\rho}=\frac n2\left(\frac r{r_c}\right)^n
  \longrightarrow 0\qquad(r\to0).
\end{equation}
Thus the calibrated completion has an exactly isotropic center. Together with the gauge identity $p_r=-\rho$, this gives the de~Sitter core expected of a regular black hole in this effective reconstruction. Away from the center the source becomes anisotropic, as expected for an effective regular interior. This does not prove uniqueness or DRG compatibility. The inverse source has the desired de~Sitter form, while the independently prescribed DRG pressure fails the consistency check in the main text.

\subsection{Toy evolution using Eulerian finite volumes}
\label{subsec:app_eulerfv}

For an exploratory check of boundedness, a collapse problem is solved using Eulerian finite volumes on a fixed grid. The DRG scalar sector enters algebraically. The code uses a gravity source of Newtonian form with the pressure correction of the TOV equation,
\begin{equation}
  g_{\rm eff} \propto
  -\rho\,
  \frac{m + 4\pi r^3 p_{\rm grav}}{r^2}
  \left(1 - \frac{\rho_{\rm tot}}{\rho_{\rm crit}}\right),
\end{equation}
with $\rho_{\rm crit} = (1 + A/2)\,\rho_P$ in the imposed bounce factor, includes the DRG scalar pressure in the momentum flux and evolves compact Gaussian initial data on a uniform grid with $N=400$ out to $t=100$. For numerical control it clips the velocity at $|v|\leq0.5$, uses von Neumann-Richtmyer viscosity with $C_Q=2$ and takes $p_{\rm grav}=\max(p_{\rm tot},0)$ in the gravity source. These interventions are not consequences of the covariant DRG equations. The sign of $f(r)=1-2m(r)/r$ serves only as an indicator of trapped regions, not as a trapped-surface calculation in a solved relativistic evolution.

\begin{figure}[t]
  \centering
  \includegraphics[width=0.98\textwidth]{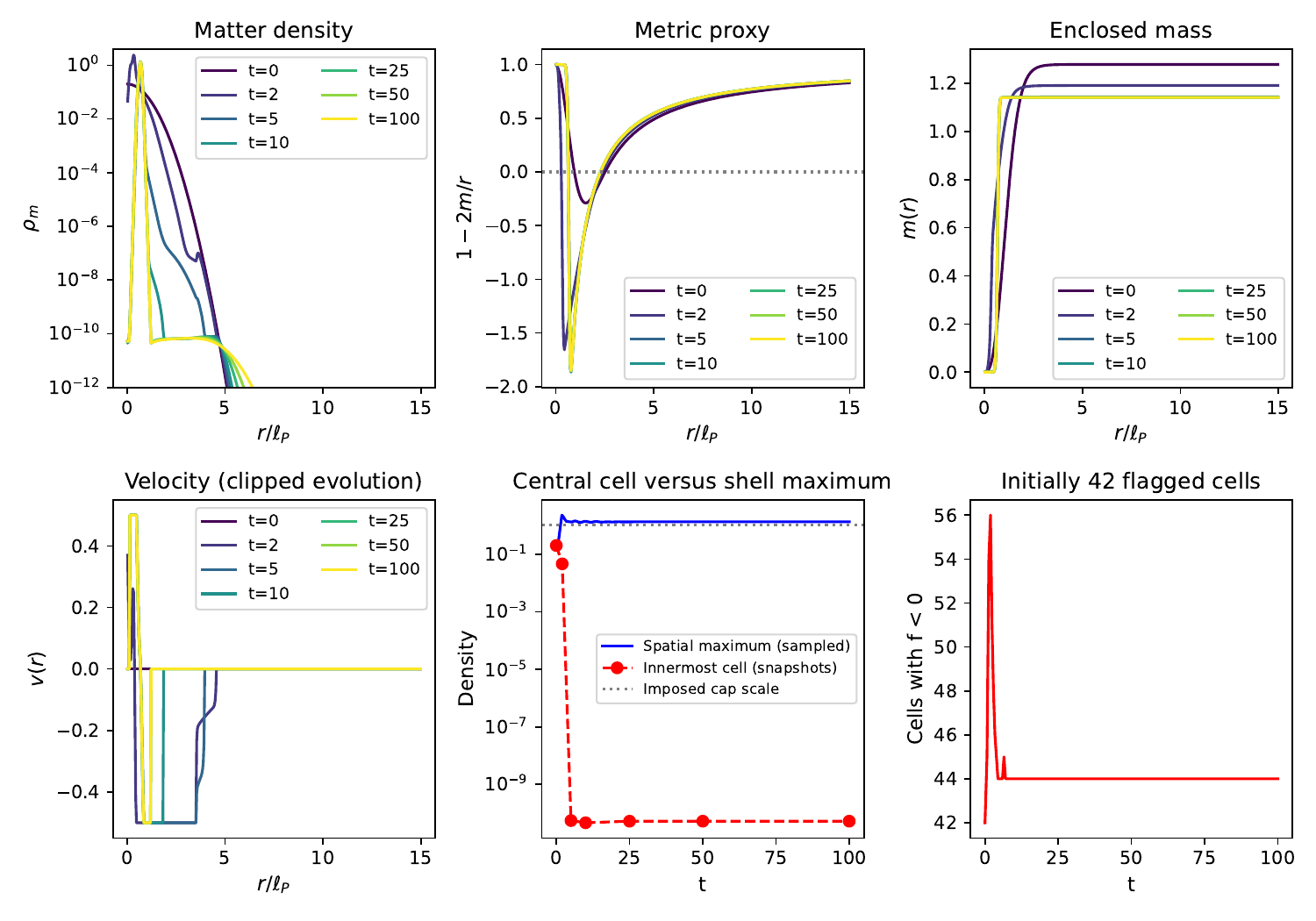}
  \caption{Six-panel summary of the $\rho_c=0.20\,\rho_P$ Eulerian
  toy evolution with an imposed bounce factor, clipping and viscosity. The gray line marks the scale $\rho_{\rm crit}$ of the bounce factor, not a density ceiling.
  Top row: density profile, $f(r)=1-2m(r)/r$ and mass function at selected times.
  Bottom row: the full clipped velocity range, a comparison of the sampled spatial maximum of the density (blue) with the density of the innermost cell at the output times (red) and the number of cells with $f<0$. These cells already exist in
  the initial data. The red density markers are values at the output times, not a resolved central time series. This figure documents the toy
  evolution and is not a relativistic bounce or formation result.}
  \label{fig:euler_fv}
\end{figure}

In a scan over nine central densities from $\rho_c = 0.02$ to $0.20\,\rho_P$, the two lowest-density runs remain subcritical, while the seven runs with $\rho_c\geq0.04\,\rho_P$ have cells with $f<0$ and remain bounded in density through $t=100$. Figure~\ref{fig:euler_fv} summarizes the selected run with $\rho_c=0.20\,\rho_P$, which already has 42 such cells at the initial time. Its spatial density maximum settles to $1.3263\,\rho_P$, whereas the density in the innermost cell falls to $5.30\times10^{-11}\,\rho_P$. The material at high density is in a shell, not at the center. The bounce flag of the code detects only a decline of the sampled spatial maximum. It is not used as evidence of a relativistic bounce. These clipped, viscous calculations establish neither the formation of a regular black hole nor conservation and stability of a covariant DRG evolution.

\section{Validation and reproducibility}
\label{app:validation}

\subsection{Asymptotic GR limit}
\label{subsec:app_gr}

As a consistency check, the relation between mass and temperature must recover the Schwarzschild result $T_H = 1/(8\pi M)$ for $M \gg \MPl$. In the limit of large mass ($r_h \gg \ell$), the parametric relations of the Hayward-type control expand as
\begin{equation}
  M(r_h)
  = \frac{r_h}{2}\!\left(1 + \frac{\ell^3}{r_h^3}\right)
  \;\Rightarrow\; r_h \approx 2M\,,
\end{equation}
\begin{equation}
  T_H(r_h)
  = \frac{1}{4\pi\,r_h}\,
    \frac{1 - 2(\ell/r_h)^3}{1 + (\ell/r_h)^3}
  \approx \frac{1}{4\pi\,r_h}\,.
\end{equation}
Substituting $r_h \approx 2M$ gives $T_H \approx 1/(8\pi M)$, as expected. This agreement is confirmed numerically in Figure~\ref{fig:gr_diagnostic}.

\subsection{Reproducibility}
\label{subsec:reproducibility}

Table~\ref{tab:reproducibility} lists all parameters and settings needed to reproduce the results of this work. The companion code and numerical outputs are available from the author upon reasonable request, as stated in the data availability section.

\begin{table}[!t]
  \caption{Parameters and settings for reproducibility.
  All quantities are in Planck units
  ($G = c = \hbar = k_B = 1$) unless stated otherwise.}
  \label{tab:reproducibility}
  \begin{ruledtabular}
  \begin{tabular}{lll}
    \multicolumn{3}{c}{\textit{Historical diagnostic inputs}} \\
    Coupling & $A$ & $0.024$ \\
    Source exponent & $\gamma$ & $0.501$ \\
    Extremal mass & $\Mmin^{\rm bench}$ & $\sqrt{1+A/2}\,\MPl = 1.00598\,\MPl$ \\[4pt]
    \multicolumn{3}{c}{\textit{Smooth DRG black hole}} \\
    Central matter density & $\rho_c$ & $0.1560\,\rho_P$ \\
    Central total density & $\rho_{\rm tot}(0)$ & $0.1647\,\rho_P$ \\
    Central Kretschmann & $K(0)$ & $45.7\,\MPl^4$ \\
    Pressure mismatch & $|p_{\rm DRG}-p_r|/\rho$ & $0.959$ \\
    ADM mass at extremality & $M_{\rm ext}$ & $1.0060\,\MPl$ \\[4pt]
    \multicolumn{3}{c}{\textit{Calibrated $n=2$ effective metric family motivated by DRG}} \\
    Calibrated $\rho_0$ & & $0.1531\,\rho_P$ \\
    Extremal radius & $r_\star$ & $1.597\,\lPl$ \\
    Remnant entropy & $\Srem$ & $8.01$ nats \\[4pt]
    \multicolumn{3}{c}{\textit{Control metrics}} \\
    Hayward core scale & $\ell$ & $1.065\,\lPl$ \\
    Bardeen core scale & $g$ & $0.774\,\lPl$ \\[4pt]
    \multicolumn{3}{c}{\textit{TOV stellar branch}} \\
    Polytropic equation of state & $K,\;\Gamma$ & $0.1,\;2.0$ \\
    Density scan & $\rho_c$ & $[10^{-6},\,10]\,\rho_P$ (250 pts) \\
    Surface threshold & $\rho_{\rm surf}$ & $10^{-8}\,\rho_P$ \\[4pt]
    \multicolumn{3}{c}{\textit{Black hole branch}} \\
    Temperature grid & $M/\MPl$ & $[1.000001\,\Mmin/\MPl,\,50\,\Mmin/\MPl]$ (450 pts) \\
    Horizon solver & & Brent on each side of the metric minimum \\
    Horizon check & & $|f(r_h)|<2\times10^{-12}$ \\[4pt]
    \multicolumn{3}{c}{\textit{Evaporation}} \\
    Blackbody coefficient & $\sigma$ & $\pi^2/60$ \\
    Greybody factor & $\mathcal G$ & $1$ (illustrative) \\
    Initial mass & $M_0$ & $5\,\MPl$ \\[4pt]
    \multicolumn{3}{c}{\textit{Crossing of the area entropy}} \\
    Initial masses tested & $M_0/\MPl$ & $5,\,10,\,20,\,50$ \\
    Families tested & & $n = 2,3,4,6$, Hayward, Bardeen \\
    Result for $n=2$ & $M_{\rm cross}/M_0$ & $0.707107$ \\
    Controls at $M_0=5\,\MPl$ & Hayward-type, Bardeen & $0.708671$, $0.713555$ \\[4pt]
    \multicolumn{3}{c}{\textit{Static diagnostic based on coordinate time}} \\
    Reference cutoff & $\epsilon$ & $10^{-6}$ \\
    Cutoff range & & $10^{-3}$ to $10^{-8}$ \\
    \multicolumn{3}{c}{\textit{Exploratory trial entropy}} \\
    Baseline & $\alpha_A,c_{\rm eff},r_{\rm bath}/\lPl$ & $1,1,100$ \\
    Separate $f$ regulator & $\epsilon_f$ & $10^{-8}$ (scan $10^{-6}$ to $10^{-10}$) \\
  \end{tabular}
  \end{ruledtabular}
\end{table}

\subsection{Exploratory functional inspired by QES}
\label{subsec:app_qes_saddles}

\begin{figure}[t]
  \centering
  \includegraphics[width=0.98\textwidth]{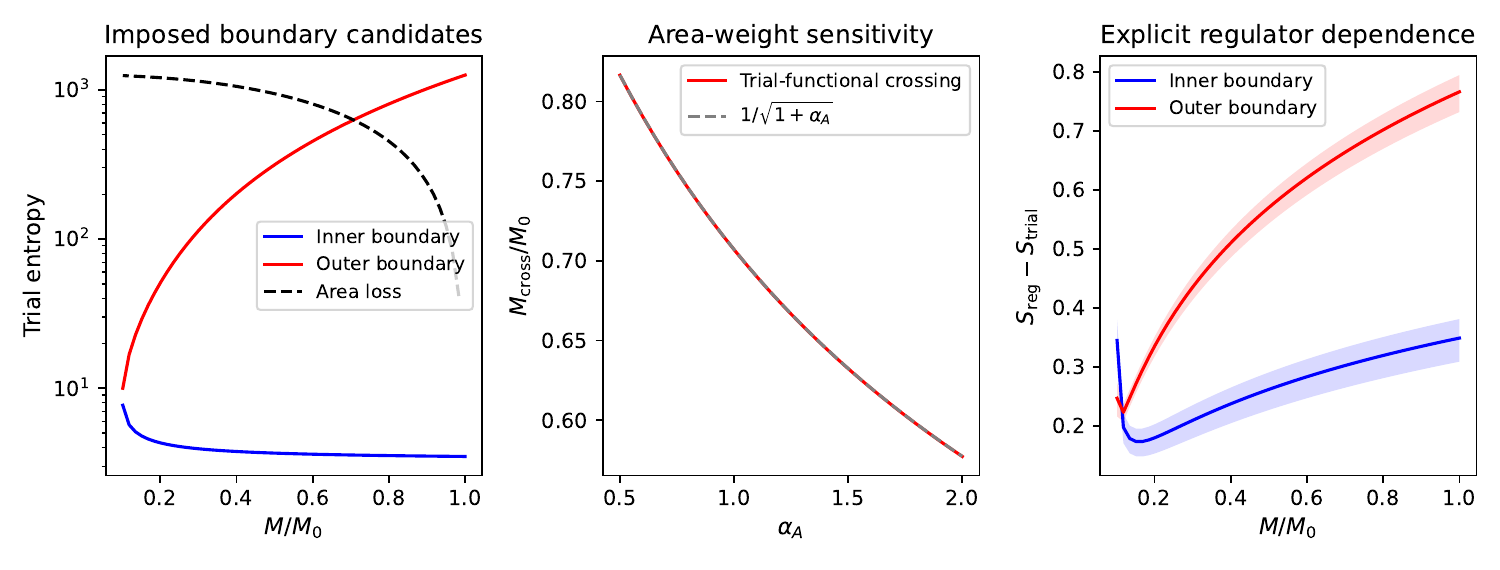}
  \caption{Exploratory sensitivity analysis of the functional inspired by QES on the calibrated DRG $n=2$
  family.
  Left: trial entropy at the two imposed boundaries and geometric
  area loss. Middle: crossing at the outer boundary versus the area weight, with its limit dominated by the area term as a dashed curve.
  Right: $S_{\rm reg}-S_{\rm trial}$ for the two boundaries,
  with lines for $\epsilon_f=10^{-8}$ and shaded ranges for
  $10^{-6}$ to $10^{-10}$. The correction is shown directly rather
  than hidden on the much larger scale set by the area entropy.
  No physical island saddle is inferred.}
  \label{fig:qes_saddles}
\end{figure}

The sensitivity study evaluates an imposed functional inspired by QES on the calibrated DRG $n=2$ family (Figure~\ref{fig:qes_saddles}). The bulk term and candidate surfaces are not derived from a quantum state. Here $c_{\rm eff}$ is an imposed coefficient of the bulk logarithm, $\alpha_A$ multiplies the area term and $r_{\rm bath}$ is the chosen outer reference radius. These are sensitivity parameters, not quantities derived from DRG. The radius of the trial surface denoted by $a$ is unrelated to a cosmological scale factor. For $M_0=10\,\MPl$, the baseline parameters are $\alpha_A=c_{\rm eff}=1$ and $r_{\rm bath}=100\,\lPl$. In Planck units the imposed functional is
\begin{equation}
 S_{\rm trial}(a)=\alpha_A\pi a^2+
 \frac{c_{\rm eff}}6\ln\frac{r_{\rm bath}}a,
 \qquad r_-\leq a\leq r_+.
\end{equation}
Its unconstrained stationary point, $a_{\rm anal}=\sqrt{c_{\rm eff}/(12\pi\alpha_A)}$, lies below $r_-$ for the baseline. The minimum on the imposed interval is the inner boundary, not an extremal surface obtained from a state calculation. The crossing at the outer boundary with the geometric area loss occurs at $M_{\rm cross}/M_0=0.707015$.

Replacing the logarithm by a distance based on $\int_a^{r_{\rm bath}}dr/|f|$ would give a divergent result at the horizons. A separate diagnostic with an explicit regulator is defined instead,
\begin{equation}
 D_{\epsilon_f}(a,r_{\rm bath})=
 \int_a^{r_{\rm bath}}\frac{dr}{\max(|f(r)|,\epsilon_f)},
 \qquad
 S_{\rm reg}(a)=\alpha_A\pi a^2+
 \frac{c_{\rm eff}}6\ln\frac{D_{\epsilon_f}}{\lPl}.
\end{equation}
The baseline is $\epsilon_f=10^{-8}$. This is a cutoff on $|f|$, not the fractional cutoff of the horizon gap in Section~\ref{sec:echoes}. At each horizon, the region $|f|<\epsilon_f$ is resolved explicitly. The remainder is integrated using the logarithm of the distance from the horizon. For the outer boundary at $M=10\,\MPl$, the change $S_{\rm reg}-S_{\rm trial}$ is $0.76624$ at the baseline. The change is $0.73201$ and $0.79462$ for $\epsilon_f=10^{-6}$ and $10^{-10}$, respectively. These finite numbers depend on the imposed regulator. No finite unregulated correction or physical QES is inferred.

The benchmark was also scanned over $c_{\rm eff} = 0.1$ to $10$, $r_{\rm bath} = 30$ to $1000\,\lPl$ (always outside $r_+$) and $\alpha_A = 0.5$ to $2$. The outer crossing is essentially insensitive to $c_{\rm eff}$ and $r_{\rm bath}$, while $\alpha_A$ shifts the crossing as expected. The ratio $M_{\rm cross}/M_0$ ranges from approximately $0.816$ at $\alpha_A = 0.5$ to approximately $0.577$ at $\alpha_A = 2$. This sensitivity belongs to the imposed functional, not to a derived quantum entropy.

\clearpage
\bibliographystyle{apsrev4-2}
\bibliography{main}

\end{document}